\documentclass[aip,
 amsmath,amssymb,
 reprint,%
]{revtex4-1}

\usepackage{graphicx}
\usepackage{color}
\usepackage{bm}
\usepackage{amsmath}
\usepackage{amssymb}
\usepackage{hyperref}
\usepackage{float}
\usepackage[latin1]{inputenc}

\def\xcm3{\mbox{cm}^{-3}}

\def\Wxcm2{\mbox{Wcm}^{-2}}

\def\Axm2{\mbox{Am}^{-2}}

\definecolor{red}{rgb}{1,0,0}
\definecolor{blue}{rgb}{0,0,1}

\begin{document}

\author{M. E. Dieckmann}
\email[]{mark.e.dieckmann@liu.se}\affiliation{Department of Science and Technology, Link\"oping University, SE-60174 Norrk\"oping, Sweden}

\author{S. J. Spencer}
\affiliation{Centre for Fusion, Space and Astrophysics, University of Warwick, Coventry, CV4 7AL, UK}

\author{M. Falk}
\affiliation{Department of Science and Technology, Link\"oping University, SE-60174 Norrk\"oping, Sweden}

\author{G. Rowlands}
\affiliation{Centre for Fusion, Space and Astrophysics, University of Warwick, Coventry, CV4 7AL, UK}

\date{\today}

\title{Preferential acceleration of positrons by a filamentation instability between an electron-proton beam and a pair plasma beam}

\begin{abstract}
Particle-in-cell (PIC) simulations of collisionless jets of electrons and positrons in an ambient electron-proton plasma have revealed an acceleration of positrons at the expense of electron kinetic energy. The dominant instability within the jet was a filamentation instability between electrons, protons and positrons. In this work we show that a filamentation instability, between an initially unmagnetized ambient electron-proton plasma at rest and a beam of pair plasma that moves through it at a non-relativistic speed, indeed results in preferential positron acceleration. Filaments form that are filled predominantly with particles with the same direction of their electric current vector. Positron filaments are separated by electromagnetic fields from beam electron filaments. Some particles can cross the field boundary and enter the filament of the other species. Positron filaments can neutralize their net charge by collecting the electrons of the ambient plasma while protons cannot easily follow the beam electron filaments. Positron filaments can thus be compressed to a higher density and temperature than the beam electron filaments. Filament mergers, which take place after the exponential growth phase of the instability has ended, lead to an expansion of the beam electron filaments, which amplifies the magnetic field they generate and induces an electric field in this filament. Beam electrons lose a substantial fraction of their kinetic energy to the electric field. Some positrons in the beam electron filament are accelerated by the induced electric field to almost twice their initial speed. The simulations show that a weaker electric field is induced in the positron filament and particles in this filament hardly change their speed.
\end{abstract}

\maketitle

\section{Introduction}
Annihilation radiation from the bulge of our galaxy evidences a sizeable positron population.~\cite{Positron1,Positron2} The huge mean free path of the interstellar medium means that positrons can travel far without annihilating, \cite{Jean2009} and it is therefore difficult to determine their source. A 2012 review of the potential explanations for the measured positron flux \cite{Serpico2012} concluded that the simplest explanation is that there exists a population of primary positron accelerators. Microquasar jets are one candidate for these primary accelerators.  Microquasars \cite{Microquasar1} are binary systems in which a compact object, typically a black hole, accretes material from a companion star. A fraction of the gravitational energy, released by the inward-falling material, powers a pair of jets, which can reach relativistic speeds and can contain positrons.~\cite{Mirabel92,Microquasar2} Microquasar jets expand into an ambient plasma, which is initially coronal plasma,~\cite{Yuan2014} followed by the companion's stellar wind and eventually the interstellar medium. A significant fraction of the positrons must be able to escape from jets if they were to be an important source of the galactic positron population. The present work aims at identifying processes that could turn a jet into such a source. 

We focus on positron acceleration mechanisms in collisionless plasma. In such a plasma, the interaction between a beam of electrons and positrons and ambient plasma composed of electrons and protons can lead to the formation of leptonic jets with a structure that resembles that of hydrodynamic jets.~\cite{Bromberg2011} Previous work has shown that positrons gained energy at the expense of the beam electron energies. This energy redistribution was observed for jets in unmagnetized plasmas \cite{Jet1} and in plasma that was permeated by a magnetic field parallel to the jet's expansion direction.~\cite{Jet2} The exact reason for this redistribution has eluded us. 

Why is it important to identify the mechanisms responsible for a preferential acceleration of positrons? A dilute cloud of energetic positrons could escape in the previous jet simulations without accompanying beam electrons. The large energy gap between the escaping positrons and the cool electrons of the ambient plasma ahead of the jet reduces the probability for their annihilation. The mechanism, that accelerates the jet's positrons at the expense of the energy of its electrons, could thus turn a jet into an effective source of energetic positrons.

The simplest model for a pair-plasma microquasar jet expanding into an ambient plasma is a beam of electrons and positrons that flows across an unmagnetized ambient electron-proton plasma. This situation might be realistic if we consider the plasma at the jet's front also known as its head once it has left the strongly magnetized coronal plasma near the accretion disk. Magnetic fields in stellar winds and in the interstellar medium tend to have an energy density that is small compared to the kinetic energy density of a relativistic jet as we discuss below.

In previous studies, letting a pair beam stream across an unmagnetized ambient electron-proton plasma~\cite{Beam1} drove an instability between the electrons of both plasmas. Geometric simulation constraints forced it to be electrostatic. Beam electrons were slowed down and positrons were accelerated. However, it was an electromagnetic filamentation instability~\cite{Davidson1972,Lee1973,Tzoufras2006} rather than an electrostatic instability that developed in the aforementioned jet simulations, in particular at the jet's head. Filamentation instabilities have been examined in pure pair plasma\cite{Filamentation1,Filamentation2} but no preferential positron acceleration was recorded. Protons can modify the filamentation instability~\cite{Baryonloading} and here we want to determine if and how they are responsible for the preferential acceleration of positrons. 

We study with the EPOCH PIC simulation code~\cite{Arber2015} the filamentation instability with a particle composition similar to the one close to the head of a jet. Positrons are accelerated by the electromagnetic fields that grow in our simulations, while beam electrons lose a large fraction of their initial speed. This energy redistribution sheds light onto why positrons gained energy at the expense of the beam electrons in the jet simulations. One intriguing consequence of observing a preferential positron acceleration in such a simple setting is that we can verify it experimentally. It is possible to create clouds of electrons and positrons in laser-plasma experiments with a particle number that is large enough to have it behave like a plasma~\cite{Gahn2002,Hui2009,Sarri2013} and to trigger instabilities.~\cite{Warwick2017}

The outline of this paper is as follows: Section \ref{Sec:setup} discusses the initial conditions for our simulations and relevant aspects of the filamentation instability. Section \ref{Sec:results} presents our simulation results and Section \ref{Sec:Discussion} summarizes them.

\section{Simulation setup}\label{Sec:setup}


An unmagnetized electron-proton plasma at rest represents the ambient plasma. Beams of electrons and positrons form the jet material. A non-relativistic beam speed simplifies the interpretation of the results because it minimizes effects due to a relativistic mass increase. The simulation box is oriented perpendicular to the beam velocity vector, which excludes all instabilities but the filamentation instability. We refer to related simulations for a discussion of electrostatic instabilities.~\cite{Beam1,Timofeev2010,Sgattoni2017}

We perform three simulations with varying sizes and dimensions of the simulation boxes: in the first, we select a simulation box that is so short that only one wave oscillation is resolved;~\cite{Beam2} we then increase the box length by a factor 5 in the second, which allows filaments to merge after they complete the linear growth phase and enter their fully nonlinear stage.~\cite{Vanthieghem2018} Our third simulation tests how a second resolved dimension affects the positron acceleration. The side length of the two-dimensional box matches that of simulation 2, which allows filaments to move around each other, to merge and grow to increasingly larger ones.~\cite{Silva2003,Medvedev2005} 
We align the one-dimensional simulation boxes with the \textit{x}-axis while the two-dimensional simulation resolves the \textit{x}-\textit{y} plane. Boundary conditions are periodic in all simulations. In simulation 1, we use 1000 grid cells to resolve the box length $L_x$. The longer box of simulation 2 is resolved by 5000 cells while we use 1000 larger cells along each direction to resolve the 2D box of simulation 3. 

Electrons of the ambient plasma have the density $n_0$ and temperature $T_0$ = 100 eV, which corresponds to the thermal speed $v_{te} \approx 0.014 c$. The thermal speed is defined as $v_{te}={(k_B T_0/m_e)}^{1/2}$ ($k_B, m_e, c$: Boltzmann constant, electron mass and speed of light). The electron plasma frequency $\omega_{pe}={(n_0e^2/\epsilon_0 m_e)}^{1/2}$ ($e, \epsilon_0, \mu_0$: elementary charge, vacuum permittivity and permeability) normalizes time and the skin depth $\lambda_s = c/\omega_{pe}$ normalizes space. If we normalize the magnetic field \textbf{B} to $m_e\omega_{pe}/e$, the electric field \textbf{E} to $m_e \omega_{pe}c/e$, the current density $\mathbf{J}$ to $en_0c$, velocities $\mathbf{v}$ to $c$ and relativistic momenta $\mathbf{p}$ to $m_ec$, the density $n_0$ drops out of the Maxwell-Lorentz force set of equations. Therefore, the numerical value of $n_0$ does not affect the plasma dynamics as long as the plasma is collisionless. It only matters when we transform the computed quantities to physical units. 

The normalized box length in simulation~1 is $L_x = 1.88$, it is $5L_x=9.4$ in simulation~2 and simulation~3 resolves an area with the size $5L_x \times 5 L_x$. The density $n_0$ and temperature $T_0$ are also given to the protons with the mass $m_p = 1836m_e$. Both species are at rest in the simulation frame. A beam of electrons and positrons moves at the mean speed $v_b$ that corresponds to the relativistic momentum $|\mathbf{p}_b| = 0.3$. The beam moves along $y$ in the 1D simulations and along $z$ in the 2D simulation. Each beam species has the number density $n_0$ and temperature $T_0$. The plasma is initially unmagnetized and free of any net charge or current. We set $\mathbf{E}$ and $\mathbf{B}$ to zero at the time $t=0$. We use 2500 (400) computational particles per cell for each species in the 1D (2D) simulations and resolve the time interval $t=90$. 

Let us estimate the amplitude of the beam-aligned magnetic field $B_{c}$ that would suppress the filamentation instability of counterstreaming electron beams.~\cite{Cary1981,Stockem2008,Bret2009,Grassi2017} We obtain $eB_{c}/m_e \omega_{pe} = v_b/c$ in physical units for our nonrelativistic beam speed $v_b$ and equal densities of the counterstreaming electron beams.~\cite{Stockem2008} Consider a plasma with the density $10 \,\mathrm{cm}^{-1}$ and magnetic field strength $5$ nT of the solar wind near the Earth's orbit. The magnetic field of the interstellar medium has a similar amplitude and we may expect that the magnetic fields of stellar winds of black hole companions are usually not much stronger. A magnetic field with the amplitude $B_c \approx 3.5 \times 10^{-7}$ T would suppress the filamentation instability; the aforementioned plasmas are thus practically unmagnetized and we are justified in using no background magnetic field in our simulations

We can understand the growth of magnetic $\mathbf{B}$ fields and electric $\mathbf{E}$ fields in response to the filamentation instability using Amp\`ere's law and Faraday's law. Under the specific initial conditions described above, and the one-dimensional simulation geometry in simulations 1 and 2, these laws can be simplified under the assumption that $\mathbf{J}=(j_x,j_y,j_z)$ of the beam electrons and positrons points along \textit{y} giving $B_y\approx 0$. The $B_x$ component cannot change in the 1D geometry due to $\nabla \cdot \mathbf{B}=0$, and hence our governing equations become:
\begin{equation}
-\frac{dB_z}{dx} = j_y + \frac{dE_y}{dt},
\label{Ampere}
\end{equation}
\begin{equation}
\frac{dB_z}{dt} = -\frac{dE_y}{dx}    
\label{Faraday}
\end{equation}

In the absence of the ambient plasma, all field fluctuations would travel with the beam and the fluctuations would not grow. However, the ambient plasma lets magnetic fluctuations move relative to the beam~\cite{Pelletier2019} and the plasma becomes unstable.

\section{Simulation results}\label{Sec:results}

\subsection{Simulation 1: Short 1D box}

\begin{figure*}[ht]
\includegraphics[width=\textwidth]{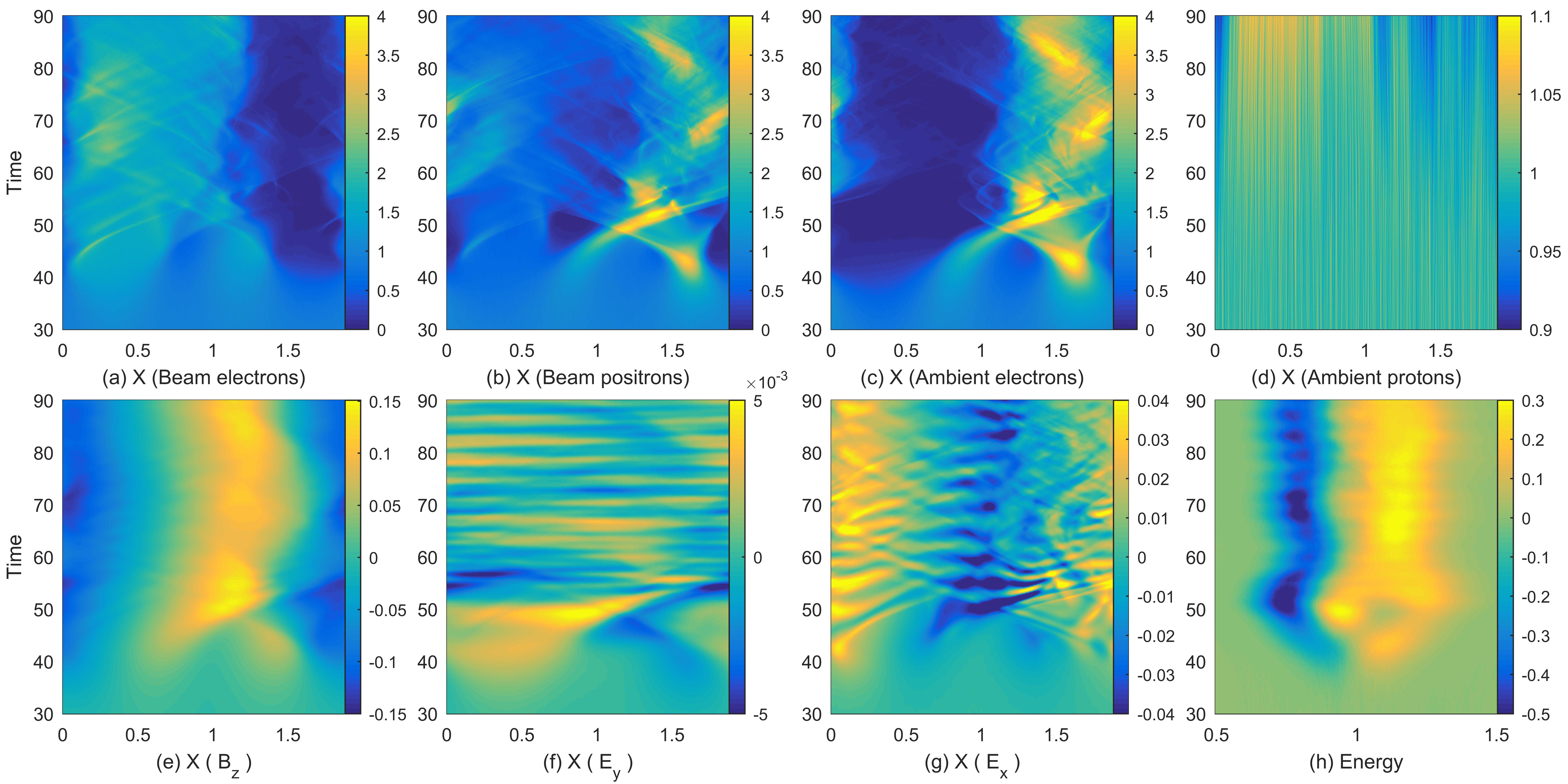}
\caption{Time evolution of the relevant field amplitudes, species densities and energy distribution in simulation 1: Panels (a-c) show the density distributions of the beam electrons, positrons and ambient electrons. Panel (d) shows the proton density. Panels (e-g) show $B_z$, $E_y$ and $E_x$, respectively. Panel (h) shows the box-averaged difference of the energy distributions of positrons and electrons $f_p(E_{kin})-f_e(E_{kin})$ with the normalized kinetic energy $E_{kin}=\mathbf{v}^2/v_b^2$ on the horizontal axis. Densities are normalized to $n_0$. Electric and magnetic field field amplitudes are normalized to $m_e\omega_{pe}c/e$ and $m_e\omega_{pe}/e$, respectively. The energy distribution is normalized to the peak value of the velocity distribution of one lepton species at $t=0$. Times $t<30$ are clipped.}
\label{figure1}
\end{figure*}

Figure~\ref{figure1} confirms that our beam configuration leads to instability. 
It reveals a redistribution of the three lepton species after $t\approx 40$ and the onset of proton density oscillations, which remain weak until $t=90$. Protons start to accumulate after $t>70$ in the interval $0.1 \le x \le 1$ where we find most beam electrons. The redistribution of the leptons correlates well with the growth and saturation of electromagnetic fields which we interpret as follows, neglecting the proton reaction. Consider a small fluctuation of $B_z$ with $dB_z/dx \approx 0$ that moves relative to the beam. It deflects the beam electrons and positrons into opposite directions, and their currents no longer cancel each other out everywhere. Amp\`ere's law (Eqn.~\ref{Ampere}) implies that an electric field $E_y$ grows if $j_y \neq 0$. This $E_y$ field changes $B_z$ according to Faraday's law (Eqn.~\ref{Faraday}). Indeed, we notice a bipolar $E_y$-field distribution at $35 \le t \le 45$ in Fig.~\ref{figure1}(f), which encloses a growing magnetic $B_z$-field in Fig.~\ref{figure1}(e). Both fields decouple when the filamentation instability saturates nonlinearly at $t \approx 45$. In this short simulation box, the magnetic field $B_z$ oscillates almost sinusoidally in space after it saturates.

Filamentation instabilities saturate by the magnetic trapping of particles.~\cite{Davidson1972} Magnetic trapping of particles by sinusoidal filamentation modes sets in when the particle bouncing frequency $\omega_B$ of particles near the bottom of the potential of the magnetowave becomes comparable to the growth rate of the instability. We obtain a bouncing frequency $\omega_B = {(kB_kv_b)}^{1/2}$ for a potential that varies sinusoidally with the normalized wavenumber $k=2\pi \lambda_s / L_x$ and is tied to a normalized magnetic field $B_k$ that oscillates in space with the same wavenumber. The values $k\approx 3.3$, $v_b=0.3$ and $B_k \approx 0.1$ (See Fig.~\ref{figure1}(e)) give $\omega_B \approx 0.3$, which matches the growth rate of the filamentation instability between unmagnetized cold electron beams of equal density for a beam speed $v_b$.~\cite{Stockem2008}

The $E_y$ field slows down the beam particles and transfers some of their kinetic energy to the magnetic $B_z$ component, which lets its energy density $\propto B_z^2$ grow. The magnetic field is not at rest in the simulation frame and its motion along $y$ induces a convective electric field $E_x$. Another contribution to $E_x$ is the space charge we get when the beam electrons and positrons start to form filaments. Figure~\ref{figure1}(g) confirms that an electric $E_x$ field is growing once the filamentation instability is nearing saturation. The mean polarization of $E_x$ is such that it confines electrons to the positron filament. Oscillations of $E_x$ and $E_y$ with a frequency $\omega_{pe}$ or period $\Delta t \approx 2\pi / \omega_{pe}$ are visible after $t\approx 45$. We show below that these oscillations are caused by the bouncing motion of particles in the wave potential, as expected for magnetic trapping.

The redistribution of the beams along $x$ results in space charge. Ambient electrons in Fig.~\ref{figure1}(c) are expelled after $t\approx 40$ from intervals where beam electrons accumulate in Figs.~\ref{figure1}(a). They gather in intervals with a positron excess (See Fig.~\ref{figure1}(b)). Once all positrons are expelled, beam electrons can only accumulate until their density matches that of the protons. Positron filaments can be compressed more since ambient electrons follow them and cancel out their charge.~\cite{Filamentation2} The width of the beam electron filament with a density $\approx 2$ in Fig.~\ref{figure1}(a) is of the order 1, while the positrons are compressed in an interval of width $0.7$ and reach a higher density. Accumulations of positrons and ambient electrons reach densities up to $4$ in Figs.~\ref{figure1}(b, c). These density peaks are correlated with electromagnetic field oscillations with a large amplitude and a short wavelength; they are not stationary in time.

The electric field $E_y$ is polarized such that it slows down the positrons and beam electrons in their respective filaments and transfers the freed energy to $B_z$. In principle, the current density $j_y$ should decrease until it establishes an equilibrium with the magnetic $B_z$ component and $E_y$. Figure~\ref{figure1} demonstrates that this equilibrium is not stationary in time; the values of $B_z$ and $E_y$ oscillate around their equilibrium distributions. The current density $j_y$ can be decreased either by a slowdown of the beam particles or by a reaction of the ambient plasma to $E_y$. Beam electrons push ambient electrons into the positron filament, which results in a different composition of the ambient plasma in the positron and beam electron filaments. Protons will hardly react to $E_y$ during the short simulation time. We expect that the beam electrons will slow down significantly to reduce their current density $j_y$. Ambient electrons react easily to electric fields. Their current density can cancel out some of that of the positron beam. The positron beam does not need to slow down as much as the electron beam in order to reduce its current density. The presence of ambient electrons may thus let beam electrons lose more energy than positrons. At the same time, the electromagnetic field may not be able to separate completely the beam electrons from the positrons. Beam electrons entering a positron filament will be accelerated by the electric field of the positron filament and vice versa. 

How will the asymmetry between the filaments of the electron beam and positron beam affect lepton energies? We box-average the energy distribution of the beam electrons and positrons giving $f_e(E_{kin})$ and $f_p (E_{kin})$ with $E_{kin}=\mathbf{v}^2/v_{b}^2$ ($\mathbf{v}$: velocity of the computational particles). Figure~\ref{figure1}(h) shows $(f_p(E_{kin})-f_e(E_{kin}))$, which we normalized to the peak value of $f_p(E_{kin})$ at the time $t=0$. We can clearly see that the onset of $E_y$ slows down the beam electrons more than the positrons.  

We can identify unambiguously one potential preferential acceleration mechanism by looking at the phase space density distributions of the electrons and positrons. Since we use a nonrelativistic beam speed, the distributions in the $x,p_x$ plane are decoupled from those in the $x,p_y$ plane ($p_x,p_y$: momenta in the $x$ and $y$ directions). We can consider them separately.

\begin{figure*}
\includegraphics[width=\textwidth]{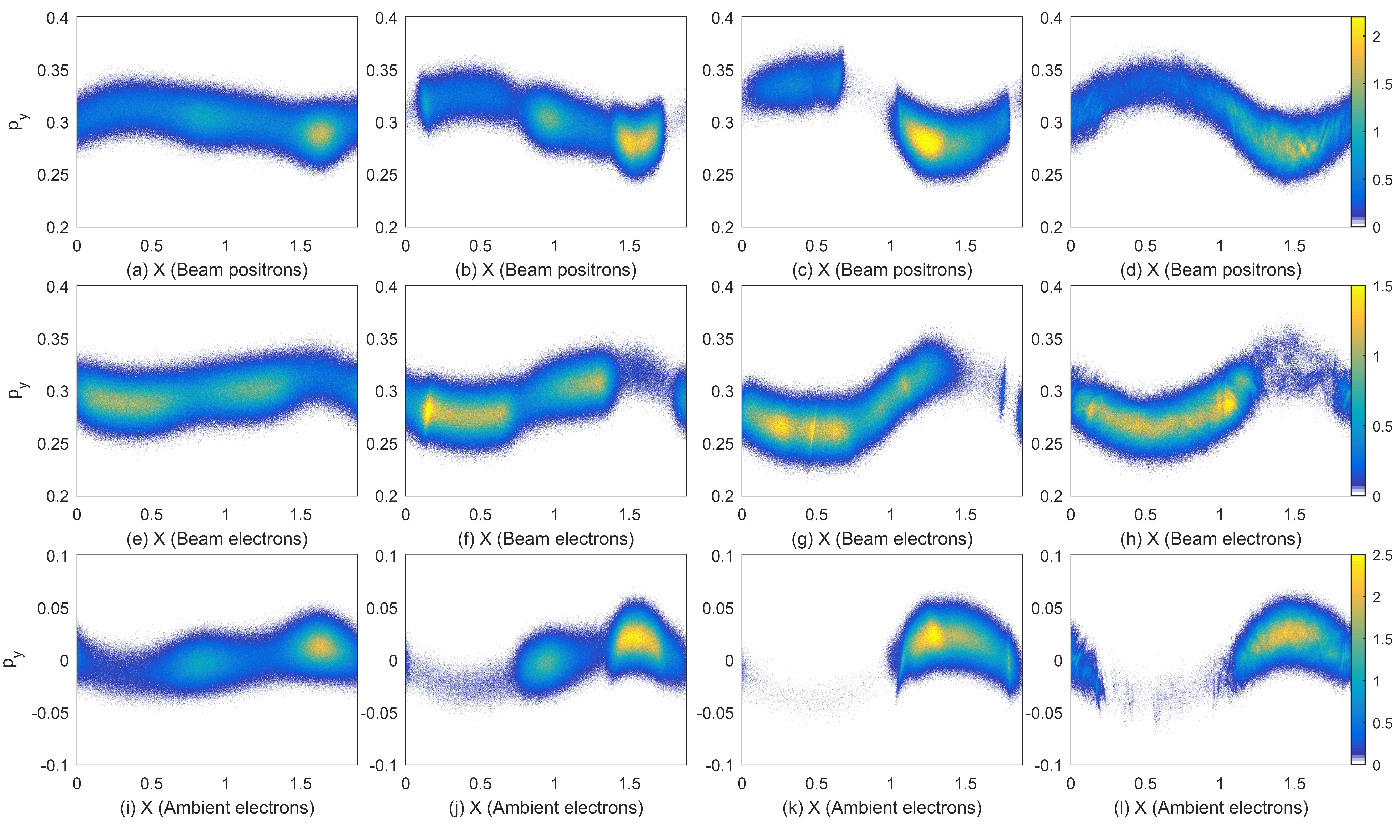}
\caption{Phase space density distributions of positrons, beam electrons, and ambient electrons in the $x,p_y$ plane at four different times in simulation 1. Panels (a, e, i) show the distributions of positrons, beam electrons and ambient electrons at the time 40. Panels (b, f, j) show the corresponding phase space density distributions at the time 45. The distributions of positrons, beam electrons and ambient electrons at the time 50 are shown in panels (c, g, k). Their distribution at the final simulation time are displayed in panels (d, h, l). All distributions are normalized to their maximum value at $t=0$ and displayed on a linear color scale. The color bar to the right is valid for all plots in the row. Multimedia view:}
\label{figure2}
\end{figure*}

Figure~\ref{figure2} depicts the projections of the phase space densities of the three lepton species onto the $x,p_y$ plane at four times.  
The left-most column (Figures~\ref{figure2}(a, e, i)) shows the distributions of positrons, beam electrons and ambient electrons at the time $t=40$, when the instability is still in its exponential growth phase. A dense positron filament has developed at $x\approx 1.6$, which is matched by a filament in the ambient electrons. As expected, the electric field $E_y$ has slowed down the positrons in this filament and accelerated the background electrons. The peak density of the beam electrons is smaller and their mean velocity changes at a lower rate over a larger $x$-interval. The ambient electrons have been diluted in the interval $0.1 \le x \le 0.5$ at $t=40$ and almost completely expelled at $t=45$ in Fig.~\ref{figure2}(j). At this time, hardly any beam electrons are left in the interval $1.4 \le x \le 1.8$ where the densest positron filament is located in Fig.~\ref{figure2}(b). The positrons have been expelled to a lesser degree from the interval $0 \le x \le 0.7$ occupied by the densest beam electron filament. At $t=50$ (Figures~\ref{figure2}(c,g,k)), most positrons and all ambient electrons have accumulated in the interval $1 \le x \le 1.8$. A diluted positron beam is still located in Fig.~\ref{figure2}(c) at $x<0.6$. Practically all beam electrons in Fig.~\ref{figure2}(g) have been expelled by the positron filament.

Figure~\ref{figure2}(Multimedia view) demonstrates that the density distributions are not stationary in time. Their motion yields the oscillations of $E_y$ in Fig.~\ref{figure1}. Figure~\ref{figure2} demonstrates that positrons and ambient electrons form a compact high-density beam at $t=90$ which expelled the beam electrons. Only a few beam electrons are thus accelerated by the electric field of the positron filament. Positrons can apparently enter the beam electron filament more easily and are accelerated by its electric field. We note that the beam particles reach their velocity minimum in the center of their respective filament because here their current density $j_y$ and the modulus $|E_y|$ of their associated electric field are largest. This explains also why the oppositely charged ambient electrons reach their largest speed in the center of the positron filament, and why positrons reach the largest speed in the center of the electron beam filament. Figure~\ref{figure2} does not, however, reveal why more positrons can enter the electron beam filament than the other way round. The phase space density distribution in the $x,p_x$ plane will reveal why this is the case. We note that the contribution of the ambient electrons to $j_y$ is negligibly small in the positron filament. Their contribution is probably not responsible for the lesser energy loss of positrons in our simulation and we must look for a different reason.

\begin{figure*}
\includegraphics[width=\textwidth]{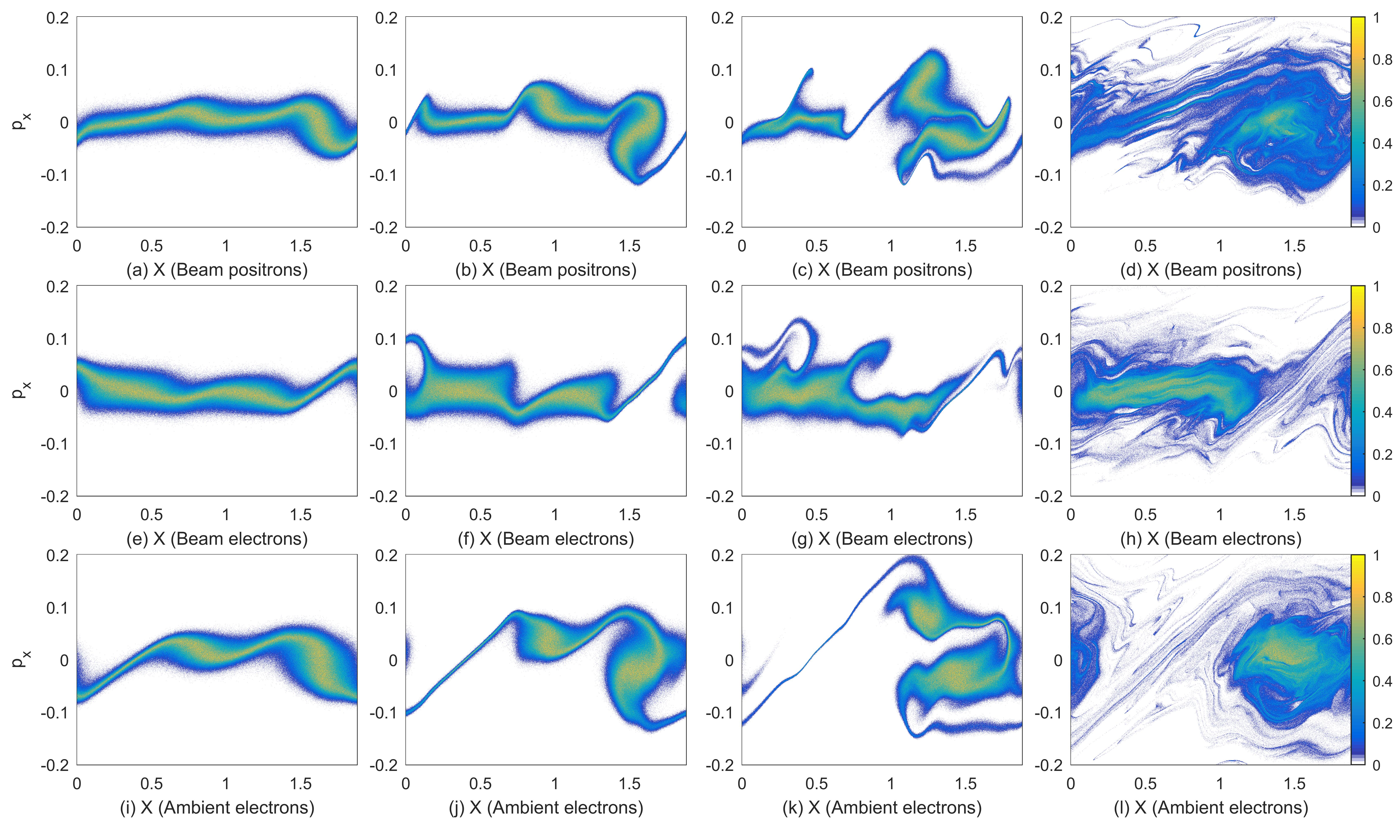}
\caption{Phase space density distributions of positrons, beam electrons, and ambient electrons in the $x,p_x$ plane at four different times in simulation 1. Panels (a, e, i) show the distributions of positrons, beam electrons and ambient electrons at the time 40. Panels (b, f, j) show the corresponding phase space density distributions at the time 45. The distributions of positrons, beam electrons and ambient electrons at the time 50 are shown in panels (c, g, k). Their distribution at the final simulation time are displayed in panels (d, h, l). All distributions are normalized to their maximum value at $t=0$ and displayed on a linear color scale. The color bar to the right is valid for all plots in the row. Multimedia view:}
\label{figure3}
\end{figure*}

Figure~\ref{figure3} shows the distributions of the positrons, beam electrons and ambient electrons at the same times as Fig.~\ref{figure2} in the phase space plane $x,p_x$.
The electric $E_x$ field and the magnetic $v_yB_z$ force accelerate the particles of all species along $x$. Some particles reach a sizable fraction of the beam's mean momentum $p_y \approx 0.3$. Figure~\ref{figure3}(Multimedia view) shows that the particles gyrate in this phase space plane, which evidences particle trapping by the electric and magnetic forces. Spots with an increased phase space density in Fig.~\ref{figure2} correspond to a distribution in Fig.~\ref{figure3} that extends over a wider velocity range and can be multivalued. 

The distribution in Fig.~\ref{figure3} is close to an equilibrium at $t=90$; positrons and ambient electrons form vortices centered at $x\approx 1.5$ with a velocity half-width of $v_b/3$. Beam electrons are distributed over a smaller velocity interval. The larger compression, which is possible for filaments of positrons and ambient electrons, heats them up to a high temperature. Beam electrons cannot be compressed much and their peak temperature is lower. We note that the particles are compressed and heated by the $v_yB_z$ force and by the electric field $E_x$ rather than by particle collisions. Their velocity distribution is thus not a Maxwellian. This electromagnetic compression implies that we find more fast positrons than beam electrons. The fastest positrons can overcome the electric and magnetic forces and enter the electron beam filament. We note in this context that the velocity of the positrons in Fig.~\ref{figure3}(d) goes through 0 in the center of the electron beam filament at $x\approx 0.5$ in Fig.~\ref{figure3}(h). Despite them having the same temperature, fewer ambient electrons overcome the electromagnetic fields and enter the beam electron filament. This asymmetry is not surprising. The presence of the protons implies that the rest frame of the growing waves is not moving at $v_b/2$. Positrons and ambient electrons thus experience different $v_yB_z$ forces.  

\subsection{Simulation 2: Long 1D box}

The short box in simulation 1 fitted only one wavelength of the unstable waves. The wave remained approximately sinusoidal after its saturation, which allowed us to determine that it saturated by magnetic trapping of leptons. This state is, however, not the final state of the instability.~\cite{Vanthieghem2018} The purpose of this simulation is to identify consequences of a continued growth of current filaments beyond the wavelength $L_x$ and of their interplay. We produce figures that can be compared directly to those presented in the previous subsection. 

\begin{figure*}[ht]
\includegraphics[width=\textwidth]{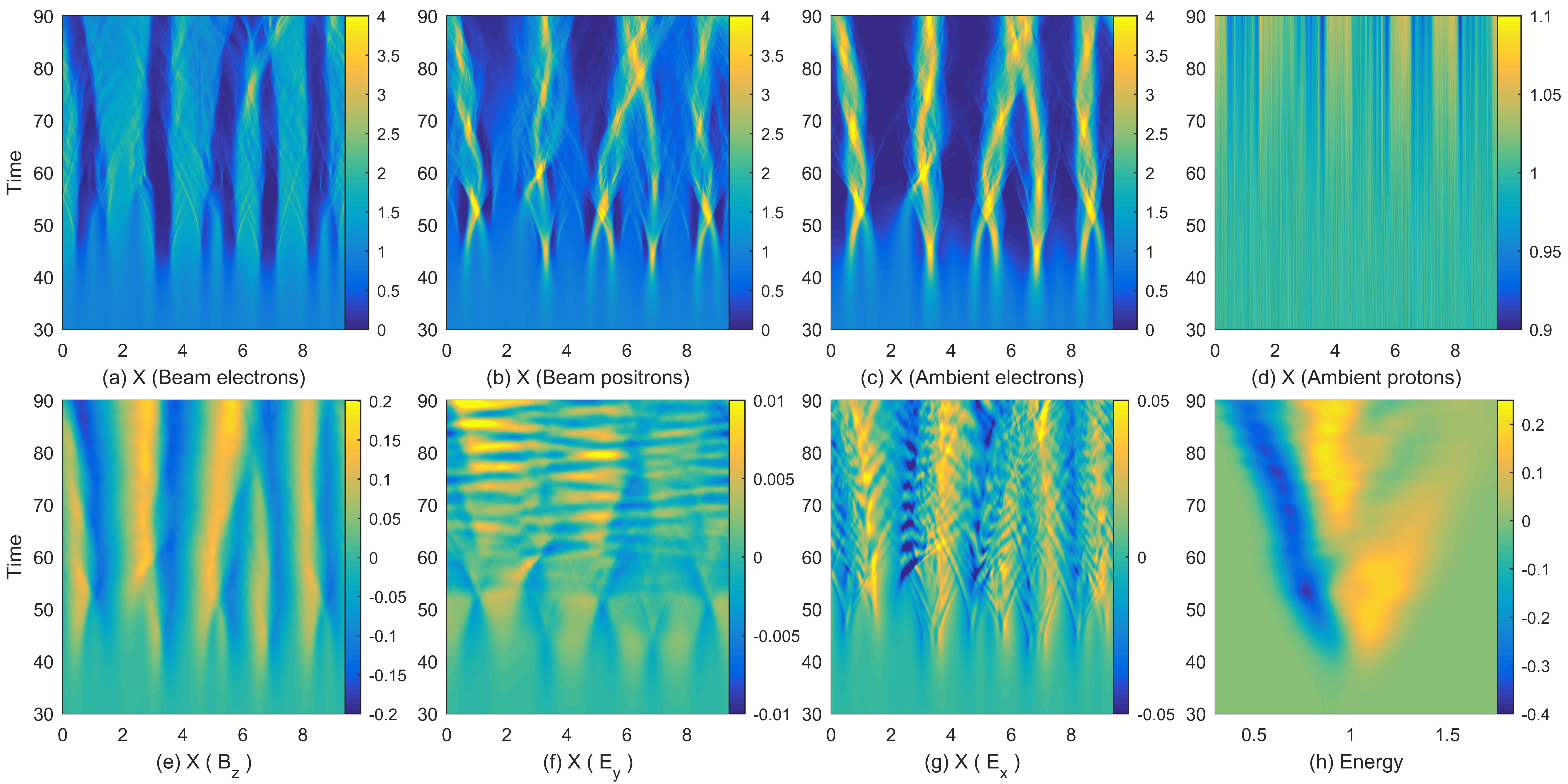}
\caption{Time evolution of the relevant field amplitudes, species densities and energy distribution in simulation 2: Panels (a-c) show the density distributions of the beam electrons, positrons and ambient electrons. Panel (d) shows the proton density. Panels (e-g) show $B_z$, $E_y$ and $E_x$, respectively. Panel (h) shows the box-averaged difference of the energy distributions of positrons and electrons $f_p(E_{kin})-f_e(E_{kin})$ with the normalized kinetic energy $E_{kin}=\mathbf{v}^2/v_b^2$ on the horizontal axis. Densities are normalized to $n_0$. Electric and magnetic field field amplitudes are normalized to $m_e\omega_{pe}c/e$ and $m_e\omega_{pe}/e$, respectively. The energy distribution is normalized to the peak value of the velocity distribution of one lepton species at $t=0$. Times $t<30$ are clipped.}
\label{figure4}
\end{figure*}
Figures~\ref{figure4}(a-c) demonstrate that 5 filaments have developed at early times $t\approx 50$. At least initially, each of these filaments has the same size as those modeled in simulation~1. Proton density oscillations remain weak but the figure confirms that protons start to accumulate in electron beam filaments. A merger between two positron filaments and their associated ambient electron filaments takes place at $x\approx 6$ and $t\approx 75$ in Figs.~\ref{figure4}(b, c). Consequently, one electron beam filament is evanescing in Fig.~\ref{figure4}(a) at $x\approx 6$ and $t\approx 80$. Its collapse frees up space for a redistribution of the other filaments. Figures~\ref{figure4}(a-c) show that the positron filaments maintain their diameter while the electron beam filaments centered at $x\approx 2$ and $x\approx 4.2$ expand in space after $t\approx 60$. The structures in the three field components displayed in Figs.~\ref{figure4}(e-g) follow closely those in the lepton density distributions. After the saturation of the magnetic $B_z$-field at $t\approx 50$, the electric $E_y$ component oscillates around a positive mean value in the intervals where the beam electron filaments are located. The oscillation amplitude of $E_y$ is twice as high as the one we observed in simulation 1 while the peak amplitude modulus $|B_z|$ exceeds that in simulation~1 by about 30\%. Strong electric $E_x$ fields ensheath the beam positron filaments. Their polarity attracts ambient electrons to the positron filaments as we already observed in Fig.~\ref{figure1}. Figure~\ref{figure4}(h) reveals a stronger energy loss of both species than in simulation 1 with beam electrons losing most of it. 

\begin{figure*}
\includegraphics[width=\textwidth]{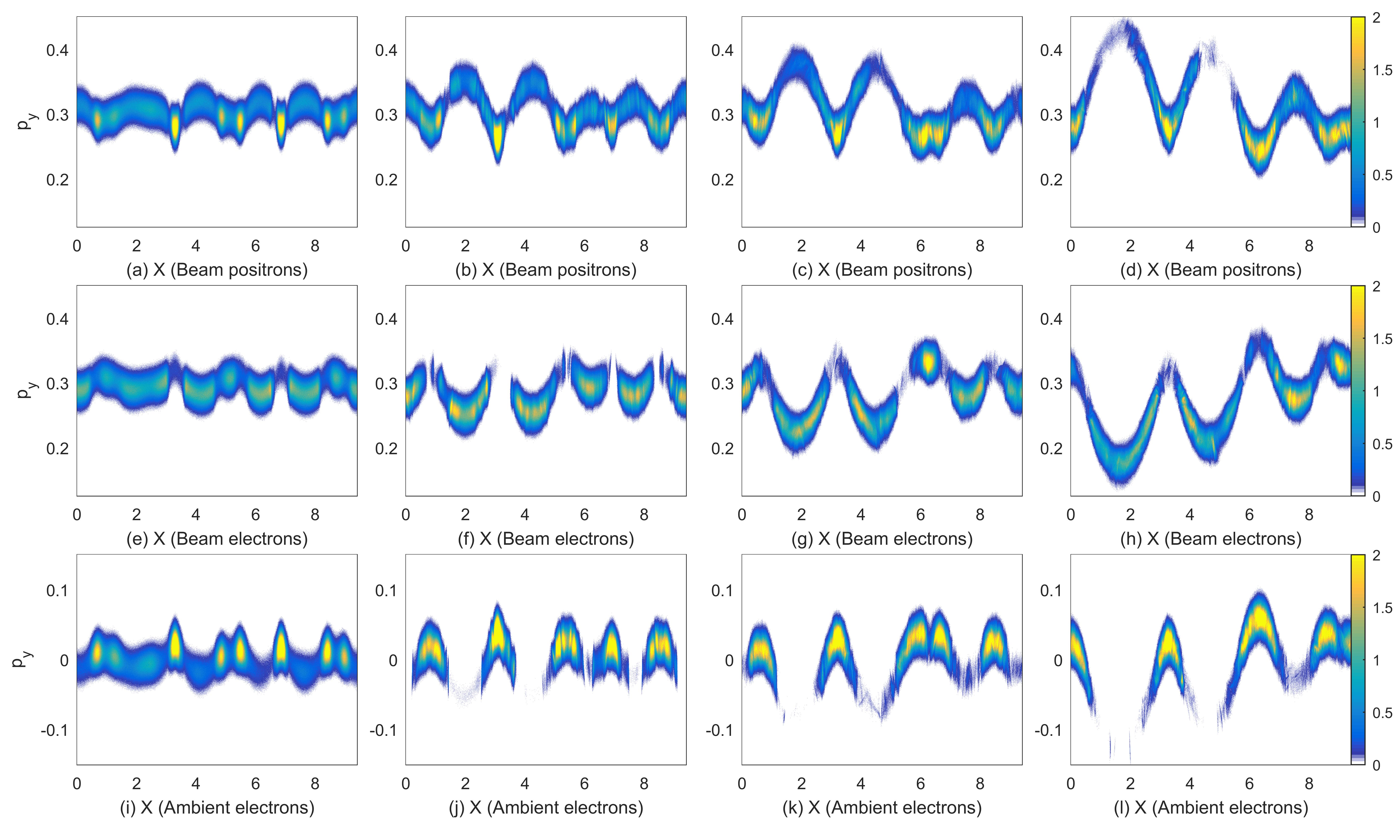}
\caption{Phase space density distributions of positrons, beam electrons, and ambient electrons in the $x,p_y$ plane at four different times in simulation 2. Panels (a, e, i) show the distributions of positrons, beam electrons and ambient electrons at the time 45. Panels (b, f, j) show the corresponding phase space density distributions at the time 60. The distributions of positrons, beam electrons and ambient electrons at the time 75 are shown in panels (c, g, k). Their distribution at the final simulation time are displayed in panels (d, h, l). All distributions are normalized to their maximum value at $t=0$ and displayed on a linear color scale. The color bar to the right is valid for all plots in the row. Multimedia view:}
\label{figure5}
\end{figure*}

How are the leptons distributed in phase space? Figure~\ref{figure5} depicts the lepton distributions in the phase space plane $x,p_y$ at 4 times.   
At the time $t=45$ when the filamentation instability saturates, the positrons and the ambient electrons form filaments with a spatial width less than 0.5. The space between the filaments is occupied by the beam electron filaments. At this time momentum changes remain small. Figures~\ref{figure5}(b, f, j) evidence a spatial separation of ambient and beam electrons. Positrons are again accelerated to a high speed inside the electron beam filaments. The particle distribution resembles at this time the one at the end of simulation~1. Figures~\ref{figure5}(c, g, k) show that the positrons have lost almost no energy in their filaments. Positrons in the electron beam filaments have gained speed along $y$ while the beam electrons have been slowed down. Beam electrons in Fig.~\ref{figure5} have suffered a dramatic speed loss at $t=90$ while the fastest positrons have almost doubled their initial speed. This is in particular true for the two electron beam filaments in the interval $0 \le x \le 4$ that expanded in Fig.~\ref{figure4}(a).

\begin{figure*}
\includegraphics[width=\textwidth]{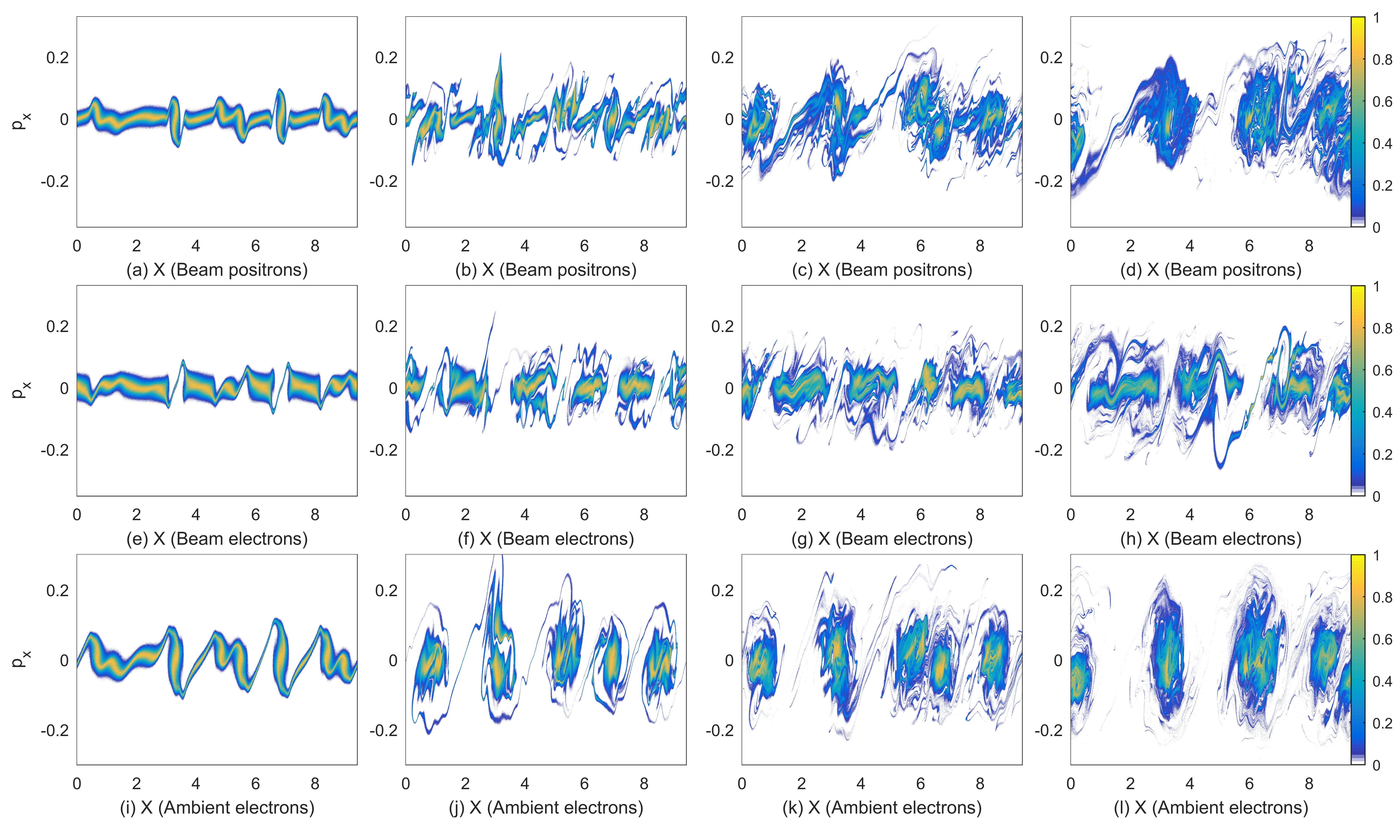}
\caption{Phase space density distributions of positrons, beam electrons, and ambient electrons in the $x,p_x$ plane at four different times in simulation 2. Panels (a, e, i) show the distributions of positrons, beam electrons and ambient electrons at the time 45. Panels (b, f, j) show the corresponding phase space density distributions at the time 60. The distributions of positrons, beam electrons and ambient electrons at the time 75 are shown in panels (c, g, k). Their distribution at the final simulation time are displayed in panels (d, h, l). All distributions are normalized to their maximum value at $t=0$ and displayed on a linear color scale. The color bar to the right is valid for all plots in the row. Multimedia view:
}
\label{figure6}
\end{figure*}

Phase space projections onto the $x,p_x$ plane at the corresponding times are shown in Fig.~\ref{figure6}.
Beam electrons have been expelled from the positron filaments at the time $t=45$. We observe a rotation of the ambient electrons and beam positrons in this plane. Figures~\ref{figure6}(b, f, j) demonstrates that filaments have fully developed at $t=60$. Positrons can still cross the electron beam filament in significant numbers while hardly any ambient electron can. Figures~\ref{figure6}(c, k) show the onset of the merger between two positron filaments at $x\approx 6$. A small electron beam filament is still present at this position in Fig.~\ref{figure6}(g). Only 4 positron filaments are left at $t=90$. Ambient electrons have formed almost circular elliptic structures, which they cannot leave. The large positron filaments at $x<4$ in Fig.~\ref{figure6}(d) can still exchange positrons. Positrons moving between both filaments are accelerated by the electric $E_y$ field of the electron beam filament. The positron acceleration mechanism identified in simulation 1 is thus also at work in this simulation.

\begin{figure*}
\includegraphics[width=\textwidth]{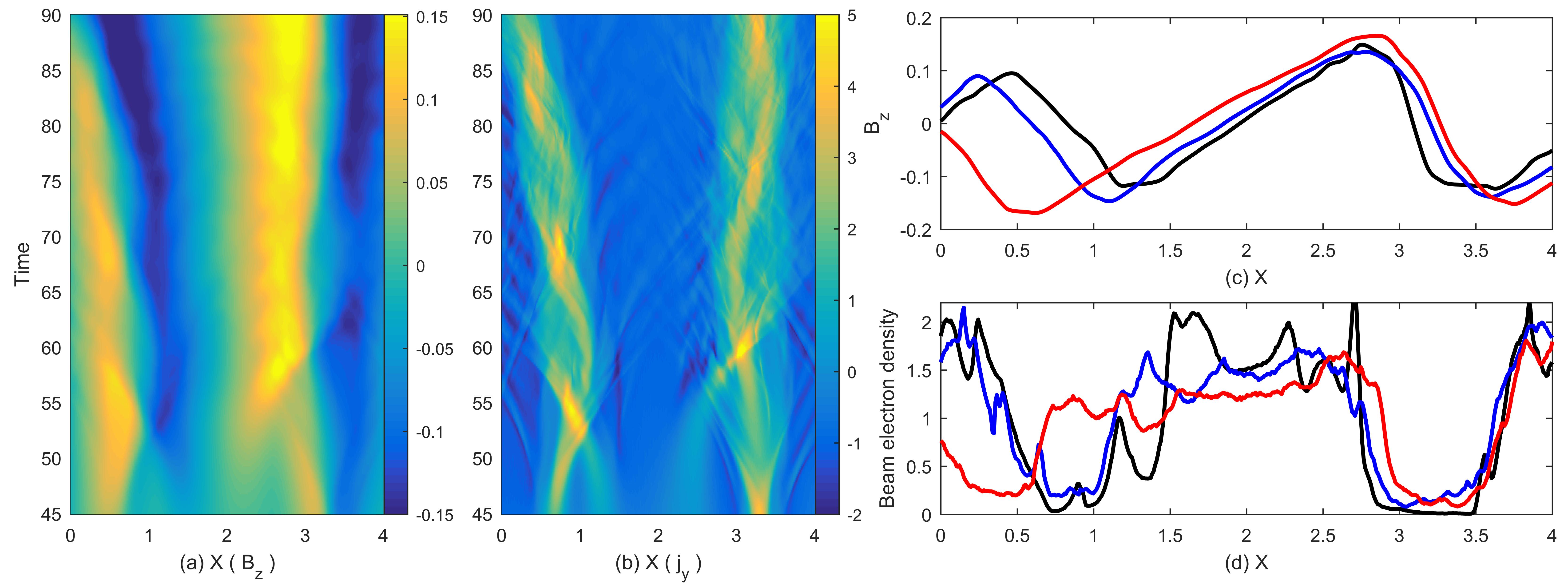}
\caption{Filament expansion in a subsection of the simulation box in simulation 2: Panel (a) shows the magnetic field $B_z$ normalized to $m_e\omega_{pe}/e$. The normalized current density $j_y/en_0c$ is displayed in panel (b). Panel (c) shows the amplitude distributions of $eB_z/m_e\omega_{pe}$ at the times $t=60$ (black), $t=75$ (blue) and $t=90$ (red). Panel (d) shows the electron beam density distribution at the same times using the same colors as panel (c).}
\label{figure7}
\end{figure*}

What is the reason for the stronger positron acceleration? Figure~\ref{figure7} provides us with a clue. It shows the distributions of the magnetic $B_z$ field, the current density component $j_y$ and slices of the electron beam density distribution at selected times. It focuses on the spatial interval where the expanding electron beam filaments are located.
Figures~\ref{figure7}(a, b) reveal a large current density in the positron filaments with the mean value $\approx 2$ reaching at times peak values of about 5. Their width is about 1. This current is responsible for a rapid decrease of $B_z$ with increasing $x$. Electron beam filaments carry less current; its modulus is about one third of the positronic current. This current is sustained over an interval with the width 1.5 at $t\approx 55$, which increases to over 2 at $t=90$. The current density carried by the electron beam filament hardly changes during this time. Figure~\ref{figure7}(c) shows slices of $B_z$ at the times $t=$ 60, 75 and 90 when the electron beam filament is growing. The slope of the magnetic amplitude hardly changes in between $1 \le x \le 2.5$ confirming that $j_y$ does not change much. The magnetic amplitude changes during this time and the magnetic energy density associated with the electron beam filament increases. This change of $B_z$ is tied via Faraday's law to an $E_y$ field that slows down the beam electrons. The magnetic energy density and amplitude changes relatively little in the positron filament.

Figure~\ref{figure7}(d) plots the electron beam density at the same times. These density slices show that the electron beam expands in time, while its density is decreasing. It remains larger than 1 for $0.5 \le x \le 3$, which is due to the need to maintain a quasi-neutrality of the plasma. The protons hardly react to the electromagnetic fields. Beam electron densities that decrease in time imply that the positron density decreases in the electron beam filaments; their current density decreases and they cancel out less of the beam electron current. This explains why the current density in the electron beam filament hardly decreases even as the density and mean velocity of the beam electrons decrease. An expanding beam electron filament accelerates positrons and gradually expels them. Simulation~2 has thus shown that the displacement electric field, which led to a preferential positron acceleration during the initial growth of the filamentation instability, continues to accelerate positrons during the filament mergers in the nonlinear phase of the instability. Such mergers were excluded in simulation~1 by the short box.

\subsection{Simulation 3: The 2-dimensional box}

Simulation 1 demonstrated that the filamentation instability saturates via magnetic trapping of leptons and that protons start to move into spatial intervals where beam electron accumulate. Simulation 2 showed that the electric field, which drives the growth and redistribution of the magnetic field during filament mergers, results in a continuing acceleration of positrons and deceleration of beam electrons. In a 2D simulation, filaments can have a cross section that changes in time and they can move around each other.~\cite{Medvedev2005,Dieckmann2007} The purpose of simulation 3 is to determine how these additional degrees of freedom affect the energy exchange between the four plasma species. 

\begin{figure*}[ht]
\includegraphics[width=\textwidth]{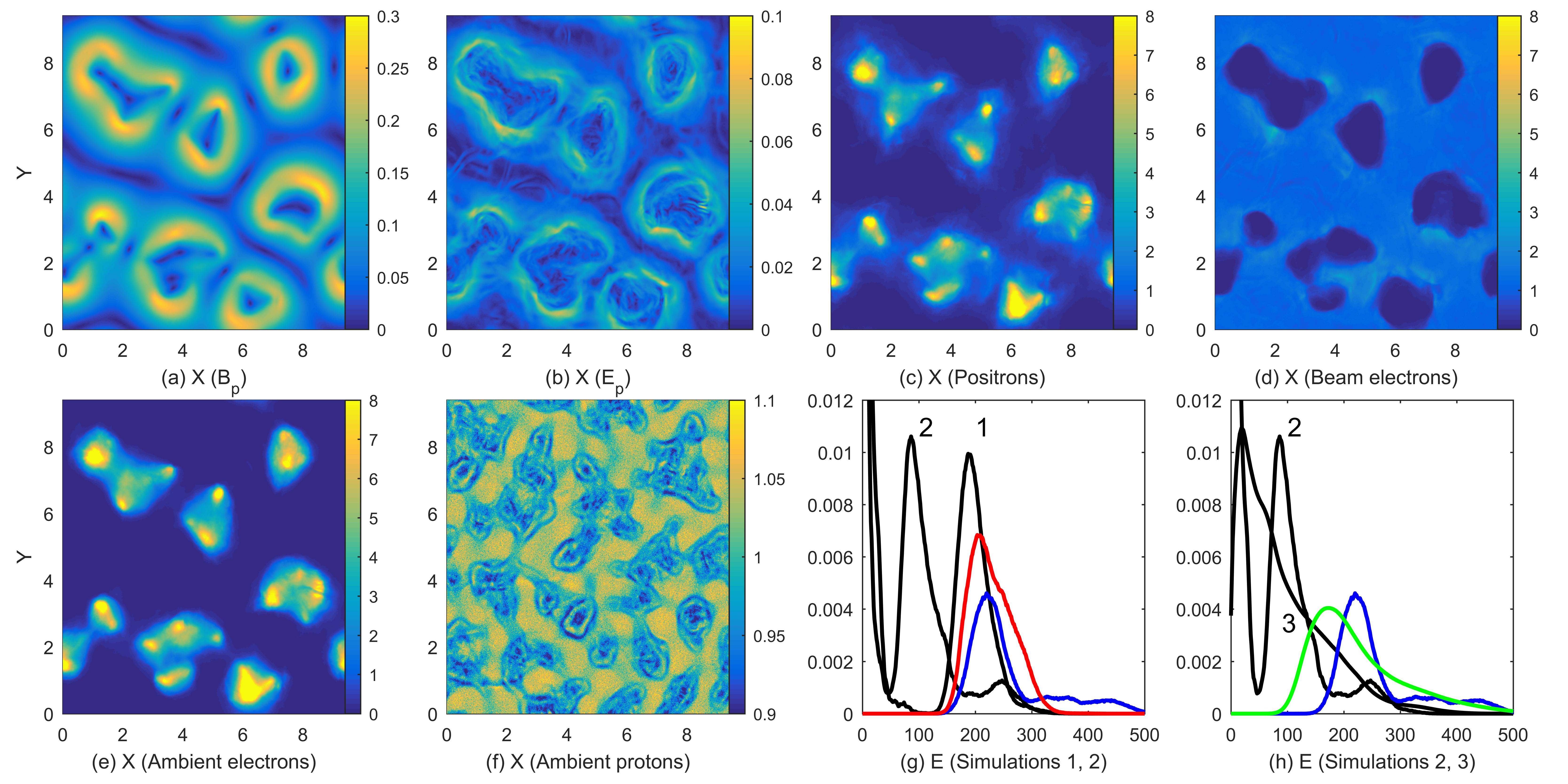}
\caption{Simulation plasma in the 2D simulation at the time $t=90$: Panel (a) and (b) show the spatial distributions of the normalized in-plane magnetic field $B_p$ and the in-plane electric field $E_p$. The positron density distribution is shown in (c) while (d) shows that of the beam electrons. Panels (e, f) present the density distributions of the ambient electrons and protons. Panels (g) and (h) compare the lepton energy distributions measured in the rest frame of the ambient plasma. Electron distributions, which are summed over both electron species, are plotted in black together with the simulation number, the positron distributions in simulations 1, 2 and 3 are plotted in red, blue and green. Energies $E$ are normalized to $E_{th}=k_BT_0$. Beam particles have on average the energy 230 at $t=0$.}
\label{figure8}
\end{figure*}

We align the beam velocity vector with the $z$-axis. The growing fields will be oriented primarily in the simulation plane spanned by $x$ and $y$. Figure~\ref{figure8} shows the normalized spatial distributions of the in-plane magnetic and electric fields $B_p = {(B_x^2+B_y^2)}^{1/2}$ and $E_p = {(E_x^2+E_y^ 2)}^{1/2}$ and the density distributions of the plasma species at the time $t=90$.
We observe annular structures in the magnetic and electric field distributions with a diameter that is about one third of the simulation box size. These field structures surround the filaments formed by the positrons and the ambient electrons. Beam electrons form a spatially uniform background outside the field structures. Their current is responsible for the slow change of the magnetic field amplitude in between the positron filaments. Structures in the proton density distribution have no obvious correlation with structures in the electromagnetic field or in the lepton density distributions. The reason is that filaments do not move during their initial exponential growth phase (See Fig.~\ref{figure4} for times $t<50$). Protons can react easily to non-oscillatory fields that are stationary in space. Filaments start to move during the nonlinear phase. Protons cannot follow their rapid motion. Figure~\ref{figure8}(f) thus shows how the filaments were distributed before they started to move and merge. 

Figure~\ref{figure8}(g) compares the energy distributions of the positrons and all electrons in simulations 1 and 2 at $t=90$. The only difference of both simulation setups is that filaments can merge to larger ones in simulation~2. Beam electrons thus lose most of their energy during filament mergers in the nonlinear phase of the instability. Positrons maintain their mean energy but those in simulation~2 can reach much higher energies due to their acceleration in the beam electron filaments. The energy spectra of electrons and positrons in simulation 3 are smoother and broader than their counterparts in simulation 2. It is evident though that beam electrons lose more energy than positrons also in a 2D geometry. 

\begin{figure*}
\includegraphics[width=\textwidth]{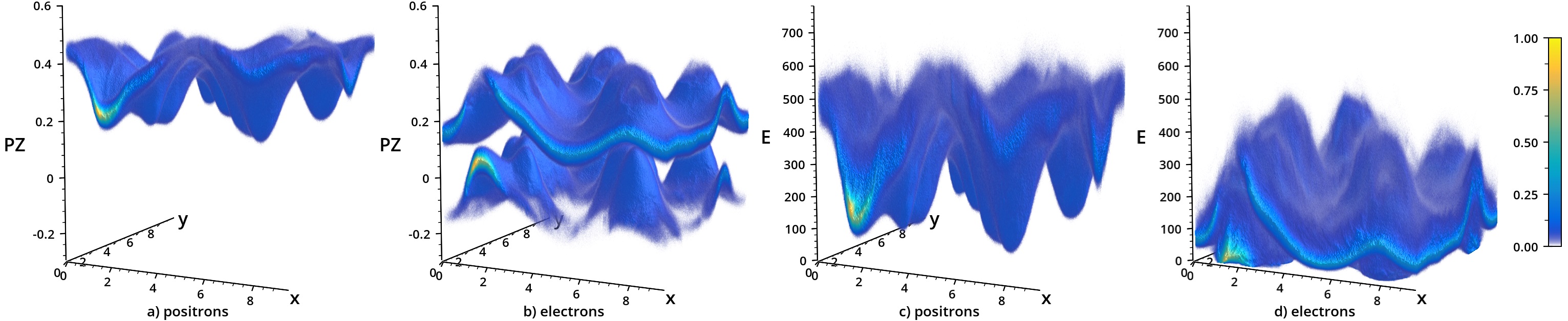}
\caption{Lepton distributions in simulation 3 at the time $t=90$: Panels (a) and (b) show the spatial distributions of the positrons and electrons as a function of the momentum component $p_z$. Panels (c) and (d) show the energy distributions of positrons and electrons as a function of space. The phase space densities in (a, b) and in (c, d) have been normalized to the largest value the electrons reached during the simulation. Color corresponds the square root of the phase space densities. The volumes have been rendered with Inviwo.~\cite{Inviwo} Energies $E$ are normalized to $E_{th}=k_BT_0$. Beam particles have on average the energy 230 at $t=0$. Multimedia view:}
\label{figure9}
\end{figure*}

Are the phase space density distributions in simulation 2 and 3 qualitatively similar? Figure~\ref{figure9} shows the projections of the phase space density distributions onto $x$, $y$ and $p_z$ and onto $x$, $y$ and the energy $E$.
Fig.~\ref{figure9}(b) demonstrates that beam electrons can enter positron filaments and be accelerated by the electric field in it. The fastest electrons can reach a momentum along $z$ that is well above their initial one. The complete expulsion of beam electrons from positron filaments in simulations 1 and 2 was thus a result of the reduced geometry. It is, however, evident from Figs.~\ref{figure9}(a, b) that beam electrons have been decelerated more than positrons in their respective filaments; beam electrons and positrons in the center of their respective filaments have mean momenta $\approx$ 0.15 and 0.2. Ambient electrons have momenta between -0.15 and 0.1. If we would slice the momentum distributions of electrons and positrons in Fig.~\ref{figure9} such that we cross filament centers, we would obtain distributions that resemble their counterparts in Figs.~\ref{figure5}(d, h, l). Figures~\ref{figure9}(c, d) demonstrate that beam electrons and positrons reach their maximum energy, measured in the rest frame of the ambient medium, in the centers of the filaments of the other beam species. Positrons reach energies that exceed those of the beam electrons by 50\%. Figure~\ref{figure9}(Multimedia view) shows the evolution of the filaments covering times between $21.5 \le t \le 90$ followed by a rotation at $t=90$. It demonstrates how filaments were initially stationary and started to move and merge when the instability entered its nonlinear phase. Their characteristic diameter just before the nonlinear phase is indeed comparable to that of the structures in the proton density distribution in Fig.~\ref{figure8}(f).

\section{Discussion}\label{Sec:Discussion}

We have examined with PIC simulations the filamentation instability between an ambient plasma consisting of protons and electrons and a pair beam. The purpose of the simulation was to determine if and how this instability can accelerate positrons to a larger speed than electrons in the rest frame of the ambient plasma. 

Our simulation showed that the current contribution of the ambient electrons in the positron filament was negligible for our initial conditions and, hence, it was not responsible for an energy loss that was larger for beam electrons than for positrons. Simulations revealed that the electrons and positrons of the pair beam reach the largest speeds in the filament formed by the other pair beam species. Filaments formed by beam positrons can reach a higher density than those formed by the beam electrons because ambient electrons can follow positrons more easily than protons can follow beam electrons. A larger compression results in a larger positron temperature when the filamentation instability saturates. Their larger temperature implies in turn that more positrons can enter an electron filament than vice versa. Hence more positrons can be accelerated by the electric field of the electron beam filament. This effect was most pronounced in the 1D simulations.
A second simulation with a larger box allowed filaments to merge, which removed one filament and enforced an expansion of the others. Positron filaments kept their size unchanged while electron beam filaments expanded. The magnetic field increased in amplitude and the displacement electric field accelerated positrons to almost twice their initial speed. It resulted in a significant slowdown of the beam electrons. The spatial expansion of the beam electron filaments, and the growth of the electromagnetic fields it caused, led to a preferential positron acceleration in the one- and two-dimensional simulations. 

The beam electron filaments attracted protons just like the positron filaments collected ambient electrons. The short simulation time implied that the proton reaction was negligible. Resolving a longer time would lead to different results in the 1D and 2D simulations. Filaments are practically stationary in 1D after they completed their mergers while those in the 2D simulation move rapidly. We expect a stronger proton reaction to the slow-moving electromagnetic fields of the filamentation instability in a one-dimensional simulation. 

Although the previous jet simulations~\cite{Jet1,Jet2} used a 2D simulation box, the filamentation instability was similar to the one-dimensional ones we considered here. This is because one simulation direction was aligned with the velocity vector of the pair beam leaving only one degree of freedom to the dynamics of the filaments. The rather generic preferential acceleration mechanism examined here may explain why positrons reached a higher energy than the electrons in our previous PIC simulations of collisionless jets. The fact that we saw a similar energy distribution also in the two-dimensional simulation~3 suggests that we would observe a preferential acceleration of positrons in three-dimensional simulations of jets.

In spite of the idealized initial conditions, the results we obtain here are important for a better understanding of positron acceleration in astrophysical jets. A jet is a pair cloud that moves at a high speed relative to the interstellar medium. Filamentation instabilities will develop all along the jet as long as we find ions in the jet that move slower than the jet material. The jet material always moves in the same direction. Our simulations revealed that on average the filamentation instability produces electric fields that point in the expansion direction of the jet. This net electric field will accelerate continuously positrons and protons in this direction. The simple plasma configuration does also lend itself to an experimental verification in laser-plasma experiments. First studies indicate that a filamentation instability can be excited in a laboratory plasma.~\cite{Warwick2017}

Future work has to examine how efficient this acceleration mechanism is for the relativistic speeds pair beams can reach in the jets of microquasars and for different ratios between the densities of the beam and the ambient plasma. It will also be interesting to determine how a beam-aligned magnetic field affects the energy redistribution between beam electrons and positrons. A magnetic field with an amplitude that can be found in the interstellar medium or in stellar winds may, however, not have a significant impact on the evolution of the instability.

\begin{acknowledgements}
The simulation was performed with the EPOCH code financed by the grant EP/P02212X/1 on resources provided by the French supercomputing facilities GENCI through the grant A0070406960 and on resources provided by the Swedish National Infrastructure for Computing (SNIC) at the HPC2N (Ume\aa ). Raw data were generated at the HPC2N (Project: SNIC2019-3-413) and GENCI (Project: A0070406960) large scale facilities. 
\end{acknowledgements}

\section*{Data availability}
Derived data supporting the findings of this study are available from the corresponding author upon reasonable request.

\bibliographystyle{aipnum4-1}
\bibliography{bib}

\begin{thebibliography}{36}%
\makeatletter
\providecommand \@ifxundefined [1]{%
 \@ifx{#1\undefined}
}%
\providecommand \@ifnum [1]{%
 \ifnum #1\expandafter \@firstoftwo
 \else \expandafter \@secondoftwo
 \fi
}%
\providecommand \@ifx [1]{%
 \ifx #1\expandafter \@firstoftwo
 \else \expandafter \@secondoftwo
 \fi
}%
\providecommand \natexlab [1]{#1}%
\providecommand \enquote  [1]{``#1''}%
\providecommand \bibnamefont  [1]{#1}%
\providecommand \bibfnamefont [1]{#1}%
\providecommand \citenamefont [1]{#1}%
\providecommand \href@noop [0]{\@secondoftwo}%
\providecommand \href [0]{\begingroup \@sanitize@url \@href}%
\providecommand \@href[1]{\@@startlink{#1}\@@href}%
\providecommand \@@href[1]{\endgroup#1\@@endlink}%
\providecommand \@sanitize@url [0]{\catcode `\\12\catcode `\$12\catcode
  `\&12\catcode `\#12\catcode `\^12\catcode `\_12\catcode `\%12\relax}%
\providecommand \@@startlink[1]{}%
\providecommand \@@endlink[0]{}%
\providecommand \url  [0]{\begingroup\@sanitize@url \@url }%
\providecommand \@url [1]{\endgroup\@href {#1}{\urlprefix }}%
\providecommand \urlprefix  [0]{URL }%
\providecommand \Eprint [0]{\href }%
\providecommand \doibase [0]{http://dx.doi.org/}%
\providecommand \selectlanguage [0]{\@gobble}%
\providecommand \bibinfo  [0]{\@secondoftwo}%
\providecommand \bibfield  [0]{\@secondoftwo}%
\providecommand \translation [1]{[#1]}%
\providecommand \BibitemOpen [0]{}%
\providecommand \bibitemStop [0]{}%
\providecommand \bibitemNoStop [0]{.\EOS\space}%
\providecommand \EOS [0]{\spacefactor3000\relax}%
\providecommand \BibitemShut  [1]{\csname bibitem#1\endcsname}%
\let\auto@bib@innerbib\@empty
\bibitem [{\citenamefont {Prantzos}\ \emph {et~al.}(2011)\citenamefont
  {Prantzos}, \citenamefont {Boehm}, \citenamefont {Bykov}, \citenamefont
  {Diehl}, \citenamefont {Ferri\`ere}, \citenamefont {Guessoum}, \citenamefont
  {Jean}, \citenamefont {Knoedlseder}, \citenamefont {Marcowith}, \citenamefont
  {Moskalenko}, \citenamefont {Strong},\ and\ \citenamefont
  {Weidenspointner}}]{Positron1}%
  \BibitemOpen
  \bibfield  {author} {\bibinfo {author} {\bibfnamefont {N.}~\bibnamefont
  {Prantzos}}, \bibinfo {author} {\bibfnamefont {C.}~\bibnamefont {Boehm}},
  \bibinfo {author} {\bibfnamefont {A.~M.}\ \bibnamefont {Bykov}}, \bibinfo
  {author} {\bibfnamefont {R.}~\bibnamefont {Diehl}}, \bibinfo {author}
  {\bibfnamefont {K.}~\bibnamefont {Ferri\`ere}}, \bibinfo {author}
  {\bibfnamefont {N.}~\bibnamefont {Guessoum}}, \bibinfo {author}
  {\bibfnamefont {P.}~\bibnamefont {Jean}}, \bibinfo {author} {\bibfnamefont
  {J.}~\bibnamefont {Knoedlseder}}, \bibinfo {author} {\bibfnamefont
  {A.}~\bibnamefont {Marcowith}}, \bibinfo {author} {\bibfnamefont {I.~V.}\
  \bibnamefont {Moskalenko}}, \bibinfo {author} {\bibfnamefont
  {A.}~\bibnamefont {Strong}}, \ and\ \bibinfo {author} {\bibfnamefont
  {G.}~\bibnamefont {Weidenspointner}},\ }\href {\doibase
  10.1103/RevModPhys.83.1001} {\bibfield  {journal} {\bibinfo  {journal} {Rev.
  Mod. Phys.}\ }\textbf {\bibinfo {volume} {83}},\ \bibinfo {pages} {1001}
  (\bibinfo {year} {2011})}\BibitemShut {NoStop}%
\bibitem [{\citenamefont {Panther}(2018)}]{Positron2}%
  \BibitemOpen
  \bibfield  {author} {\bibinfo {author} {\bibfnamefont {F.}~\bibnamefont
  {Panther}},\ }\href {\doibase 10.3390/galaxies6020039} {\bibfield  {journal}
  {\bibinfo  {journal} {Galaxies}\ }\textbf {\bibinfo {volume} {6}},\ \bibinfo
  {pages} {39} (\bibinfo {year} {2018})}\BibitemShut {NoStop}%
\bibitem [{\citenamefont {Jean}\ \emph {et~al.}(2009)\citenamefont {Jean},
  \citenamefont {Gillard}, \citenamefont {Marcowith},\ and\ \citenamefont
  {Ferri\`ere}}]{Jean2009}%
  \BibitemOpen
  \bibfield  {author} {\bibinfo {author} {\bibfnamefont {P.}~\bibnamefont
  {Jean}}, \bibinfo {author} {\bibfnamefont {W.}~\bibnamefont {Gillard}},
  \bibinfo {author} {\bibfnamefont {A.}~\bibnamefont {Marcowith}}, \ and\
  \bibinfo {author} {\bibfnamefont {K.}~\bibnamefont {Ferri\`ere}},\ }\href
  {\doibase 10.1051/0004-6361/200809830} {\bibfield  {journal} {\bibinfo
  {journal} {A\&A}\ }\textbf {\bibinfo {volume} {508}},\ \bibinfo {pages}
  {1099} (\bibinfo {year} {2009})}\BibitemShut {NoStop}%
\bibitem [{\citenamefont {Serpico}(2012)}]{Serpico2012}%
  \BibitemOpen
  \bibfield  {author} {\bibinfo {author} {\bibfnamefont {P.~D.}\ \bibnamefont
  {Serpico}},\ }\href {\doibase 10.1016/j.astropartphys.2011.08.007} {\bibfield
   {journal} {\bibinfo  {journal} {Astroparticle Physics}\ }\textbf {\bibinfo
  {volume} {39-40}},\ \bibinfo {pages} {2} (\bibinfo {year} {2012})},\ \Eprint
  {http://arxiv.org/abs/1108.4827} {arXiv:1108.4827} \BibitemShut {NoStop}%
\bibitem [{\citenamefont {Mirabel}\ and\ \citenamefont
  {Rodriguez}(1994)}]{Microquasar1}%
  \BibitemOpen
  \bibfield  {author} {\bibinfo {author} {\bibfnamefont {I.~F.}\ \bibnamefont
  {Mirabel}}\ and\ \bibinfo {author} {\bibfnamefont {L.~F.}\ \bibnamefont
  {Rodriguez}},\ }\href {https://doi.org/10.1038/371046a0} {\bibfield
  {journal} {\bibinfo  {journal} {Nature}\ }\textbf {\bibinfo {volume} {371}}
  (\bibinfo {year} {1994})}\BibitemShut {NoStop}%
\bibitem [{\citenamefont {Mirabel}\ \emph {et~al.}(1992)\citenamefont
  {Mirabel}, \citenamefont {Rodriguez}, \citenamefont {Cordier}, \citenamefont
  {Paul},\ and\ \citenamefont {Lebrun}}]{Mirabel92}%
  \BibitemOpen
  \bibfield  {author} {\bibinfo {author} {\bibfnamefont {I.~F.}\ \bibnamefont
  {Mirabel}}, \bibinfo {author} {\bibfnamefont {L.}~\bibnamefont {Rodriguez}},
  \bibinfo {author} {\bibfnamefont {B.}~\bibnamefont {Cordier}}, \bibinfo
  {author} {\bibfnamefont {J.}~\bibnamefont {Paul}}, \ and\ \bibinfo {author}
  {\bibfnamefont {F.}~\bibnamefont {Lebrun}},\ }\href
  {https://doi.org/10.1038/358215a0} {\bibfield  {journal} {\bibinfo  {journal}
  {Nature}\ }\textbf {\bibinfo {volume} {358}},\ \bibinfo {pages} {215}
  (\bibinfo {year} {1992})}\BibitemShut {NoStop}%
\bibitem [{\citenamefont {Fender}\ and\ \citenamefont
  {Gallo}(2014)}]{Microquasar2}%
  \BibitemOpen
  \bibfield  {author} {\bibinfo {author} {\bibfnamefont {R.}~\bibnamefont
  {Fender}}\ and\ \bibinfo {author} {\bibfnamefont {E.}~\bibnamefont {Gallo}},\
  }\href {\doibase 10.1007/s11214-014-0069-z} {\bibfield  {journal} {\bibinfo
  {journal} {Space Sci. Rev.}\ }\textbf {\bibinfo {volume} {183}},\ \bibinfo
  {pages} {323} (\bibinfo {year} {2014})}\BibitemShut {NoStop}%
\bibitem [{\citenamefont {Yuan}\ and\ \citenamefont
  {Narayan}(2014)}]{Yuan2014}%
  \BibitemOpen
  \bibfield  {author} {\bibinfo {author} {\bibfnamefont {F.}~\bibnamefont
  {Yuan}}\ and\ \bibinfo {author} {\bibfnamefont {R.}~\bibnamefont {Narayan}},\
  }\href {\doibase 10.1146/annurev-astro-082812-141003} {\bibfield  {journal}
  {\bibinfo  {journal} {Ann. Rev. Astron. Astrophys.}\ }\textbf {\bibinfo
  {volume} {52}},\ \bibinfo {pages} {529} (\bibinfo {year} {2014})}\BibitemShut
  {NoStop}%
\bibitem [{\citenamefont {Bromberg}\ \emph {et~al.}(2011)\citenamefont
  {Bromberg}, \citenamefont {Nakar}, \citenamefont {Piran},\ and\ \citenamefont
  {Sari}}]{Bromberg2011}%
  \BibitemOpen
  \bibfield  {author} {\bibinfo {author} {\bibfnamefont {O.}~\bibnamefont
  {Bromberg}}, \bibinfo {author} {\bibfnamefont {E.}~\bibnamefont {Nakar}},
  \bibinfo {author} {\bibfnamefont {T.}~\bibnamefont {Piran}}, \ and\ \bibinfo
  {author} {\bibfnamefont {R.}~\bibnamefont {Sari}},\ }\href {\doibase O.
  Bromberg, E. Nakar, T. Piran, and R. Sari} {\bibfield  {journal} {\bibinfo
  {journal} {Astrophys. J.}\ }\textbf {\bibinfo {volume} {740}},\ \bibinfo
  {pages} {100} (\bibinfo {year} {2011})}\BibitemShut {NoStop}%
\bibitem [{\citenamefont {Dieckmann}\ \emph
  {et~al.}(2018{\natexlab{a}})\citenamefont {Dieckmann}, \citenamefont {Sarri},
  \citenamefont {Folini}, \citenamefont {Walder},\ and\ \citenamefont
  {Borghesi}}]{Jet1}%
  \BibitemOpen
  \bibfield  {author} {\bibinfo {author} {\bibfnamefont {M.~E.}\ \bibnamefont
  {Dieckmann}}, \bibinfo {author} {\bibfnamefont {G.}~\bibnamefont {Sarri}},
  \bibinfo {author} {\bibfnamefont {D.}~\bibnamefont {Folini}}, \bibinfo
  {author} {\bibfnamefont {R.}~\bibnamefont {Walder}}, \ and\ \bibinfo {author}
  {\bibfnamefont {M.}~\bibnamefont {Borghesi}},\ }\href {\doibase
  10.1063/1.5050599} {\bibfield  {journal} {\bibinfo  {journal} {Phys.
  Plasmas}\ }\textbf {\bibinfo {volume} {25}},\ \bibinfo {pages} {112903}
  (\bibinfo {year} {2018}{\natexlab{a}})}\BibitemShut {NoStop}%
\bibitem [{\citenamefont {Dieckmann}\ \emph {et~al.}(2019)\citenamefont
  {Dieckmann}, \citenamefont {Folini}, \citenamefont {Hotz}, \citenamefont
  {Nordman}, \citenamefont {Dell\'{}Acqua}, \citenamefont {Ynnerman},\ and\
  \citenamefont {Walder}}]{Jet2}%
  \BibitemOpen
  \bibfield  {author} {\bibinfo {author} {\bibfnamefont {M.~E.}\ \bibnamefont
  {Dieckmann}}, \bibinfo {author} {\bibfnamefont {D.}~\bibnamefont {Folini}},
  \bibinfo {author} {\bibfnamefont {I.}~\bibnamefont {Hotz}}, \bibinfo {author}
  {\bibfnamefont {A.}~\bibnamefont {Nordman}}, \bibinfo {author} {\bibfnamefont
  {P.}~\bibnamefont {Dell\'{}Acqua}}, \bibinfo {author} {\bibfnamefont
  {A.}~\bibnamefont {Ynnerman}}, \ and\ \bibinfo {author} {\bibfnamefont
  {R.}~\bibnamefont {Walder}},\ }\href {\doibase 10.1051/0004-6361/201834393}
  {\bibfield  {journal} {\bibinfo  {journal} {A\&A}\ }\textbf {\bibinfo
  {volume} {621}},\ \bibinfo {pages} {A142} (\bibinfo {year}
  {2019})}\BibitemShut {NoStop}%
\bibitem [{\citenamefont {Dieckmann}\ \emph
  {et~al.}(2018{\natexlab{b}})\citenamefont {Dieckmann}, \citenamefont {Alejo},
  \citenamefont {Sarri}, \citenamefont {Folini},\ and\ \citenamefont
  {Walder}}]{Beam1}%
  \BibitemOpen
  \bibfield  {author} {\bibinfo {author} {\bibfnamefont {M.~E.}\ \bibnamefont
  {Dieckmann}}, \bibinfo {author} {\bibfnamefont {A.}~\bibnamefont {Alejo}},
  \bibinfo {author} {\bibfnamefont {G.}~\bibnamefont {Sarri}}, \bibinfo
  {author} {\bibfnamefont {D.}~\bibnamefont {Folini}}, \ and\ \bibinfo {author}
  {\bibfnamefont {R.}~\bibnamefont {Walder}},\ }\href {\doibase
  10.1063/1.5026568} {\bibfield  {journal} {\bibinfo  {journal} {Phys.
  Plasmas}\ }\textbf {\bibinfo {volume} {25}},\ \bibinfo {pages} {064502}
  (\bibinfo {year} {2018}{\natexlab{b}})}\BibitemShut {NoStop}%
\bibitem [{\citenamefont {Davidson}\ \emph {et~al.}(1972)\citenamefont
  {Davidson}, \citenamefont {Hammer}, \citenamefont {Haber},\ and\
  \citenamefont {Wagner}}]{Davidson1972}%
  \BibitemOpen
  \bibfield  {author} {\bibinfo {author} {\bibfnamefont {R.~C.}\ \bibnamefont
  {Davidson}}, \bibinfo {author} {\bibfnamefont {D.~Ã.}\ \bibnamefont
  {Hammer}}, \bibinfo {author} {\bibfnamefont {I.}~\bibnamefont {Haber}}, \
  and\ \bibinfo {author} {\bibfnamefont {C.~E.}\ \bibnamefont {Wagner}},\
  }\href {\doibase 10.1063/1.1693910} {\bibfield  {journal} {\bibinfo
  {journal} {Phys. Fluids}\ }\textbf {\bibinfo {volume} {15}},\ \bibinfo
  {pages} {317} (\bibinfo {year} {1972})}\BibitemShut {NoStop}%
\bibitem [{\citenamefont {Lee}\ and\ \citenamefont {Lampe}(1973)}]{Lee1973}%
  \BibitemOpen
  \bibfield  {author} {\bibinfo {author} {\bibfnamefont {R.}~\bibnamefont
  {Lee}}\ and\ \bibinfo {author} {\bibfnamefont {M.}~\bibnamefont {Lampe}},\
  }\href {\doibase 10.1103/PhysRevLett.31.1390} {\bibfield  {journal} {\bibinfo
   {journal} {Phys. Rev. Lett.}\ }\textbf {\bibinfo {volume} {31}},\ \bibinfo
  {pages} {1390} (\bibinfo {year} {1973})}\BibitemShut {NoStop}%
\bibitem [{\citenamefont {Tzoufras}\ \emph {et~al.}(2006)\citenamefont
  {Tzoufras}, \citenamefont {Ren}, \citenamefont {Tsung}, \citenamefont
  {Tonge}, \citenamefont {Mori}, \citenamefont {Fiore}, \citenamefont
  {Fonseca},\ and\ \citenamefont {Silva}}]{Tzoufras2006}%
  \BibitemOpen
  \bibfield  {author} {\bibinfo {author} {\bibfnamefont {M.}~\bibnamefont
  {Tzoufras}}, \bibinfo {author} {\bibfnamefont {C.}~\bibnamefont {Ren}},
  \bibinfo {author} {\bibfnamefont {F.~S.}\ \bibnamefont {Tsung}}, \bibinfo
  {author} {\bibfnamefont {J.~W.}\ \bibnamefont {Tonge}}, \bibinfo {author}
  {\bibfnamefont {W.~B.}\ \bibnamefont {Mori}}, \bibinfo {author}
  {\bibfnamefont {M.}~\bibnamefont {Fiore}}, \bibinfo {author} {\bibfnamefont
  {R.~Ã.}\ \bibnamefont {Fonseca}}, \ and\ \bibinfo {author} {\bibfnamefont
  {L.~O.}\ \bibnamefont {Silva}},\ }\href {\doibase
  10.1103/PhysRevLett.96.105002} {\bibfield  {journal} {\bibinfo  {journal}
  {Phys. Rev. Lett.}\ }\textbf {\bibinfo {volume} {96}},\ \bibinfo {pages}
  {105002} (\bibinfo {year} {2006})}\BibitemShut {NoStop}%
\bibitem [{\citenamefont {Silva}\ \emph
  {et~al.}(2003{\natexlab{a}})\citenamefont {Silva}, \citenamefont {Fonseca},
  \citenamefont {Tonge}, \citenamefont {Dawson}, \citenamefont {Mori},\ and\
  \citenamefont {Medvedev}}]{Filamentation1}%
  \BibitemOpen
  \bibfield  {author} {\bibinfo {author} {\bibfnamefont {L.~O.}\ \bibnamefont
  {Silva}}, \bibinfo {author} {\bibfnamefont {R.~A.}\ \bibnamefont {Fonseca}},
  \bibinfo {author} {\bibfnamefont {J.~W.}\ \bibnamefont {Tonge}}, \bibinfo
  {author} {\bibfnamefont {J.~M.}\ \bibnamefont {Dawson}}, \bibinfo {author}
  {\bibfnamefont {W.~B.}\ \bibnamefont {Mori}}, \ and\ \bibinfo {author}
  {\bibfnamefont {M.~V.}\ \bibnamefont {Medvedev}},\ }\href {\doibase
  10.1086/379156} {\bibfield  {journal} {\bibinfo  {journal} {Astrophys. J.}\
  }\textbf {\bibinfo {volume} {596}},\ \bibinfo {pages} {L121} (\bibinfo {year}
  {2003}{\natexlab{a}})}\BibitemShut {NoStop}%
\bibitem [{\citenamefont {Dieckmann}, \citenamefont {Shukla},\ and\
  \citenamefont {Stenflo}(2009)}]{Filamentation2}%
  \BibitemOpen
  \bibfield  {author} {\bibinfo {author} {\bibfnamefont {M.~E.}\ \bibnamefont
  {Dieckmann}}, \bibinfo {author} {\bibfnamefont {P.~K.}\ \bibnamefont
  {Shukla}}, \ and\ \bibinfo {author} {\bibfnamefont {L.}~\bibnamefont
  {Stenflo}},\ }\href {\doibase 10.1088/0741-3335/51/6/065015} {\bibfield
  {journal} {\bibinfo  {journal} {Plasma Phys. Control. Fusion}\ }\textbf
  {\bibinfo {volume} {51}},\ \bibinfo {pages} {065015} (\bibinfo {year}
  {2009})}\BibitemShut {NoStop}%
\bibitem [{\citenamefont {Fiore}\ \emph {et~al.}(2006)\citenamefont {Fiore},
  \citenamefont {Silva}, \citenamefont {Ren}, \citenamefont {Tzoufras},\ and\
  \citenamefont {Mori}}]{Baryonloading}%
  \BibitemOpen
  \bibfield  {author} {\bibinfo {author} {\bibfnamefont {M.}~\bibnamefont
  {Fiore}}, \bibinfo {author} {\bibfnamefont {L.~O.}\ \bibnamefont {Silva}},
  \bibinfo {author} {\bibfnamefont {C.}~\bibnamefont {Ren}}, \bibinfo {author}
  {\bibfnamefont {M.~A.}\ \bibnamefont {Tzoufras}}, \ and\ \bibinfo {author}
  {\bibfnamefont {W.~B.}\ \bibnamefont {Mori}},\ }\href {\doibase
  10.1111/j.1365-2966.2006.10980.x} {\bibfield  {journal} {\bibinfo  {journal}
  {MNRAS}\ }\textbf {\bibinfo {volume} {372}},\ \bibinfo {pages} {1851}
  (\bibinfo {year} {2006})}\BibitemShut {NoStop}%
\bibitem [{\citenamefont {Arber}\ \emph {et~al.}(2015)\citenamefont {Arber},
  \citenamefont {Bennett}, \citenamefont {Brady}, \citenamefont
  {Lawrence-Douglas}, \citenamefont {Ramsay}, \citenamefont {Sircombe},
  \citenamefont {Gillies}, \citenamefont {Evans}, \citenamefont {Schmitz},
  \citenamefont {Bell},\ and\ \citenamefont {Ridgers}}]{Arber2015}%
  \BibitemOpen
  \bibfield  {author} {\bibinfo {author} {\bibfnamefont {T.~D.}\ \bibnamefont
  {Arber}}, \bibinfo {author} {\bibfnamefont {K.}~\bibnamefont {Bennett}},
  \bibinfo {author} {\bibfnamefont {C.~S.}\ \bibnamefont {Brady}}, \bibinfo
  {author} {\bibfnamefont {A.}~\bibnamefont {Lawrence-Douglas}}, \bibinfo
  {author} {\bibfnamefont {M.~G.}\ \bibnamefont {Ramsay}}, \bibinfo {author}
  {\bibfnamefont {N.~J.}\ \bibnamefont {Sircombe}}, \bibinfo {author}
  {\bibfnamefont {P.}~\bibnamefont {Gillies}}, \bibinfo {author} {\bibfnamefont
  {R.~G.}\ \bibnamefont {Evans}}, \bibinfo {author} {\bibfnamefont
  {H.}~\bibnamefont {Schmitz}}, \bibinfo {author} {\bibfnamefont {A.~R.}\
  \bibnamefont {Bell}}, \ and\ \bibinfo {author} {\bibfnamefont {C.~P.}\
  \bibnamefont {Ridgers}},\ }\href {\doibase 10.1088/0741-3335/57/11/113001}
  {\bibfield  {journal} {\bibinfo  {journal} {Plasma Phys. Control. Fusion}\
  }\textbf {\bibinfo {volume} {57}},\ \bibinfo {pages} {113001} (\bibinfo
  {year} {2015})}\BibitemShut {NoStop}%
\bibitem [{\citenamefont {Gahn}\ \emph {et~al.}(2002)\citenamefont {Gahn},
  \citenamefont {Tsakiris}, \citenamefont {Pretzler}, \citenamefont {Witte},
  \citenamefont {Thirolf}, \citenamefont {Habs}, \citenamefont {Delfin},\ and\
  \citenamefont {Wahlstrom}}]{Gahn2002}%
  \BibitemOpen
  \bibfield  {author} {\bibinfo {author} {\bibfnamefont {C.}~\bibnamefont
  {Gahn}}, \bibinfo {author} {\bibfnamefont {G.~D.}\ \bibnamefont {Tsakiris}},
  \bibinfo {author} {\bibfnamefont {G.}~\bibnamefont {Pretzler}}, \bibinfo
  {author} {\bibfnamefont {K.~J.}\ \bibnamefont {Witte}}, \bibinfo {author}
  {\bibfnamefont {P.}~\bibnamefont {Thirolf}}, \bibinfo {author} {\bibfnamefont
  {D.}~\bibnamefont {Habs}}, \bibinfo {author} {\bibfnamefont {C.}~\bibnamefont
  {Delfin}}, \ and\ \bibinfo {author} {\bibfnamefont {C.~G.}\ \bibnamefont
  {Wahlstrom}},\ }\href {\doibase 10.1063/1.1446879} {\bibfield  {journal}
  {\bibinfo  {journal} {Phys. Plasmas}\ }\textbf {\bibinfo {volume} {9}},\
  \bibinfo {pages} {987} (\bibinfo {year} {2002})}\BibitemShut {NoStop}%
\bibitem [{\citenamefont {Chen}\ \emph {et~al.}(2009)\citenamefont {Chen},
  \citenamefont {Wilks}, \citenamefont {Bonlie}, \citenamefont {Liang},
  \citenamefont {Myatt}, \citenamefont {Price}, \citenamefont {Meyerhofer},\
  and\ \citenamefont {Beiersdorfer}}]{Hui2009}%
  \BibitemOpen
  \bibfield  {author} {\bibinfo {author} {\bibfnamefont {H.}~\bibnamefont
  {Chen}}, \bibinfo {author} {\bibfnamefont {S.~C.}\ \bibnamefont {Wilks}},
  \bibinfo {author} {\bibfnamefont {J.~D.}\ \bibnamefont {Bonlie}}, \bibinfo
  {author} {\bibfnamefont {E.~P.}\ \bibnamefont {Liang}}, \bibinfo {author}
  {\bibfnamefont {J.}~\bibnamefont {Myatt}}, \bibinfo {author} {\bibfnamefont
  {D.~F.}\ \bibnamefont {Price}}, \bibinfo {author} {\bibfnamefont {D.~D.}\
  \bibnamefont {Meyerhofer}}, \ and\ \bibinfo {author} {\bibfnamefont
  {P.}~\bibnamefont {Beiersdorfer}},\ }\href {\doibase
  10.1103/PhysRevLett.102.105001} {\bibfield  {journal} {\bibinfo  {journal}
  {Phys. Rev. Lett.}\ }\textbf {\bibinfo {volume} {102}},\ \bibinfo {pages}
  {105001} (\bibinfo {year} {2009})}\BibitemShut {NoStop}%
\bibitem [{\citenamefont {Sarri}\ \emph {et~al.}(2013)\citenamefont {Sarri},
  \citenamefont {Schumaker}, \citenamefont {Di~Piazza}, \citenamefont {Vargas},
  \citenamefont {Dromey}, \citenamefont {Dieckmann}, \citenamefont {Chvykov},
  \citenamefont {Maksimchuk}, \citenamefont {Yanovsky}, \citenamefont {He},
  \citenamefont {Hou}, \citenamefont {Nees}, \citenamefont {Thomas},
  \citenamefont {Keitel}, \citenamefont {Zepf},\ and\ \citenamefont
  {Krushelnick}}]{Sarri2013}%
  \BibitemOpen
  \bibfield  {author} {\bibinfo {author} {\bibfnamefont {G.}~\bibnamefont
  {Sarri}}, \bibinfo {author} {\bibfnamefont {W.}~\bibnamefont {Schumaker}},
  \bibinfo {author} {\bibfnamefont {A.}~\bibnamefont {Di~Piazza}}, \bibinfo
  {author} {\bibfnamefont {M.}~\bibnamefont {Vargas}}, \bibinfo {author}
  {\bibfnamefont {B.}~\bibnamefont {Dromey}}, \bibinfo {author} {\bibfnamefont
  {M.~E.}\ \bibnamefont {Dieckmann}}, \bibinfo {author} {\bibfnamefont
  {V.}~\bibnamefont {Chvykov}}, \bibinfo {author} {\bibfnamefont
  {A.}~\bibnamefont {Maksimchuk}}, \bibinfo {author} {\bibfnamefont
  {V.}~\bibnamefont {Yanovsky}}, \bibinfo {author} {\bibfnamefont {Z.~H.}\
  \bibnamefont {He}}, \bibinfo {author} {\bibfnamefont {B.~X.}\ \bibnamefont
  {Hou}}, \bibinfo {author} {\bibfnamefont {J.~A.}\ \bibnamefont {Nees}},
  \bibinfo {author} {\bibfnamefont {A.~G.~R.}\ \bibnamefont {Thomas}}, \bibinfo
  {author} {\bibfnamefont {C.~H.}\ \bibnamefont {Keitel}}, \bibinfo {author}
  {\bibfnamefont {M.}~\bibnamefont {Zepf}}, \ and\ \bibinfo {author}
  {\bibfnamefont {K.}~\bibnamefont {Krushelnick}},\ }\href {\doibase
  10.1103/PhysRevLett.110.255002} {\bibfield  {journal} {\bibinfo  {journal}
  {Phys. Rev. Lett.}\ }\textbf {\bibinfo {volume} {110}},\ \bibinfo {pages}
  {255002} (\bibinfo {year} {2013})}\BibitemShut {NoStop}%
\bibitem [{\citenamefont {Warwick}\ \emph {et~al.}(2017)\citenamefont
  {Warwick}, \citenamefont {Dzelzainis}, \citenamefont {Dieckmann},
  \citenamefont {Schumaker}, \citenamefont {Doria}, \citenamefont {Romagnani},
  \citenamefont {Poder}, \citenamefont {Cole}, \citenamefont {Alejo},
  \citenamefont {Yeung}, \citenamefont {Krushelnick}, \citenamefont {Mangles},
  \citenamefont {Najmudin}, \citenamefont {Reville}, \citenamefont {Samarin},
  \citenamefont {Symes}, \citenamefont {Thomas}, \citenamefont {Borghesi},\
  and\ \citenamefont {G.}}]{Warwick2017}%
  \BibitemOpen
  \bibfield  {author} {\bibinfo {author} {\bibfnamefont {J.}~\bibnamefont
  {Warwick}}, \bibinfo {author} {\bibfnamefont {T.}~\bibnamefont {Dzelzainis}},
  \bibinfo {author} {\bibfnamefont {M.~E.}\ \bibnamefont {Dieckmann}}, \bibinfo
  {author} {\bibfnamefont {W.}~\bibnamefont {Schumaker}}, \bibinfo {author}
  {\bibfnamefont {D.}~\bibnamefont {Doria}}, \bibinfo {author} {\bibfnamefont
  {L.}~\bibnamefont {Romagnani}}, \bibinfo {author} {\bibfnamefont
  {K.}~\bibnamefont {Poder}}, \bibinfo {author} {\bibfnamefont {J.~M.}\
  \bibnamefont {Cole}}, \bibinfo {author} {\bibfnamefont {A.}~\bibnamefont
  {Alejo}}, \bibinfo {author} {\bibfnamefont {M.}~\bibnamefont {Yeung}},
  \bibinfo {author} {\bibfnamefont {K.}~\bibnamefont {Krushelnick}}, \bibinfo
  {author} {\bibfnamefont {S.~P.~D.}\ \bibnamefont {Mangles}}, \bibinfo
  {author} {\bibfnamefont {Z.}~\bibnamefont {Najmudin}}, \bibinfo {author}
  {\bibfnamefont {B.}~\bibnamefont {Reville}}, \bibinfo {author} {\bibfnamefont
  {G.~M.}\ \bibnamefont {Samarin}}, \bibinfo {author} {\bibfnamefont {D.~D.}\
  \bibnamefont {Symes}}, \bibinfo {author} {\bibfnamefont {A.~G.~R.}\
  \bibnamefont {Thomas}}, \bibinfo {author} {\bibfnamefont {M.}~\bibnamefont
  {Borghesi}}, \ and\ \bibinfo {author} {\bibfnamefont {S.}~\bibnamefont
  {G.}},\ }\href {\doibase 10.1103/PhysRevLett.119.185002} {\bibfield
  {journal} {\bibinfo  {journal} {Phys. Rev. Lett.}\ }\textbf {\bibinfo
  {volume} {119}},\ \bibinfo {pages} {185002} (\bibinfo {year}
  {2017})}\BibitemShut {NoStop}%
\bibitem [{\citenamefont {Timofeev}\ and\ \citenamefont
  {Terekhov}(2010)}]{Timofeev2010}%
  \BibitemOpen
  \bibfield  {author} {\bibinfo {author} {\bibfnamefont {I.~V.}\ \bibnamefont
  {Timofeev}}\ and\ \bibinfo {author} {\bibfnamefont {A.~V.}\ \bibnamefont
  {Terekhov}},\ }\href {\doibase 10.1063/1.3474952} {\bibfield  {journal}
  {\bibinfo  {journal} {Phys. Plasmas}\ }\textbf {\bibinfo {volume} {17}},\
  \bibinfo {pages} {083111} (\bibinfo {year} {2010})}\BibitemShut {NoStop}%
\bibitem [{\citenamefont {Sgattoni}\ \emph {et~al.}(2017)\citenamefont
  {Sgattoni}, \citenamefont {Amiranoff}, \citenamefont {Briand}, \citenamefont
  {Henri}, \citenamefont {Grech},\ and\ \citenamefont
  {Riconda}}]{Sgattoni2017}%
  \BibitemOpen
  \bibfield  {author} {\bibinfo {author} {\bibfnamefont {A.}~\bibnamefont
  {Sgattoni}}, \bibinfo {author} {\bibfnamefont {F.}~\bibnamefont {Amiranoff}},
  \bibinfo {author} {\bibfnamefont {C.}~\bibnamefont {Briand}}, \bibinfo
  {author} {\bibfnamefont {P.}~\bibnamefont {Henri}}, \bibinfo {author}
  {\bibfnamefont {M.}~\bibnamefont {Grech}}, \ and\ \bibinfo {author}
  {\bibfnamefont {C.}~\bibnamefont {Riconda}},\ }\href {\doibase
  10.1063/1.4989724} {\bibfield  {journal} {\bibinfo  {journal} {Phys.
  Plasmas}\ }\textbf {\bibinfo {volume} {24}},\ \bibinfo {pages} {072103}
  (\bibinfo {year} {2017})}\BibitemShut {NoStop}%
\bibitem [{\citenamefont {Dieckmann}\ \emph {et~al.}(2009)\citenamefont
  {Dieckmann}, \citenamefont {Kourakis}, \citenamefont {Borghesi},\ and\
  \citenamefont {Rowlands}}]{Beam2}%
  \BibitemOpen
  \bibfield  {author} {\bibinfo {author} {\bibfnamefont {M.~E.}\ \bibnamefont
  {Dieckmann}}, \bibinfo {author} {\bibfnamefont {I.}~\bibnamefont {Kourakis}},
  \bibinfo {author} {\bibfnamefont {M.}~\bibnamefont {Borghesi}}, \ and\
  \bibinfo {author} {\bibfnamefont {G.}~\bibnamefont {Rowlands}},\ }\href
  {\doibase 10.1063/1.3160629} {\bibfield  {journal} {\bibinfo  {journal}
  {Phys. Plasmas}\ }\textbf {\bibinfo {volume} {16}},\ \bibinfo {pages}
  {074502} (\bibinfo {year} {2009})}\BibitemShut {NoStop}%
\bibitem [{\citenamefont {Vanthieghem}, \citenamefont {Lemoine},\ and\
  \citenamefont {Gremillet}(2018)}]{Vanthieghem2018}%
  \BibitemOpen
  \bibfield  {author} {\bibinfo {author} {\bibfnamefont {A.}~\bibnamefont
  {Vanthieghem}}, \bibinfo {author} {\bibfnamefont {M.}~\bibnamefont
  {Lemoine}}, \ and\ \bibinfo {author} {\bibfnamefont {L.}~\bibnamefont
  {Gremillet}},\ }\href {\doibase 10.1063/1.5033562} {\bibfield  {journal}
  {\bibinfo  {journal} {Phys. Plasmas}\ }\textbf {\bibinfo {volume} {25}},\
  \bibinfo {pages} {072115} (\bibinfo {year} {2018})}\BibitemShut {NoStop}%
\bibitem [{\citenamefont {Silva}\ \emph
  {et~al.}(2003{\natexlab{b}})\citenamefont {Silva}, \citenamefont {Fonseca},
  \citenamefont {Tonge}, \citenamefont {Dawson}, \citenamefont {Mori},\ and\
  \citenamefont {Medvedev}}]{Silva2003}%
  \BibitemOpen
  \bibfield  {author} {\bibinfo {author} {\bibfnamefont {L.~O.}\ \bibnamefont
  {Silva}}, \bibinfo {author} {\bibfnamefont {R.~A.}\ \bibnamefont {Fonseca}},
  \bibinfo {author} {\bibfnamefont {J.~W.}\ \bibnamefont {Tonge}}, \bibinfo
  {author} {\bibfnamefont {J.~M.}\ \bibnamefont {Dawson}}, \bibinfo {author}
  {\bibfnamefont {W.~B.}\ \bibnamefont {Mori}}, \ and\ \bibinfo {author}
  {\bibfnamefont {M.~V.}\ \bibnamefont {Medvedev}},\ }\href {\doibase
  10.1086/379156} {\bibfield  {journal} {\bibinfo  {journal} {Astrophys. J.}\
  }\textbf {\bibinfo {volume} {596}},\ \bibinfo {pages} {L121} (\bibinfo {year}
  {2003}{\natexlab{b}})}\BibitemShut {NoStop}%
\bibitem [{\citenamefont {Medvedev}\ \emph {et~al.}(2005)\citenamefont
  {Medvedev}, \citenamefont {Fiore}, \citenamefont {Fonseca}, \citenamefont
  {Silva},\ and\ \citenamefont {Mori}}]{Medvedev2005}%
  \BibitemOpen
  \bibfield  {author} {\bibinfo {author} {\bibfnamefont {M.~V.}\ \bibnamefont
  {Medvedev}}, \bibinfo {author} {\bibfnamefont {M.}~\bibnamefont {Fiore}},
  \bibinfo {author} {\bibfnamefont {R.~A.}\ \bibnamefont {Fonseca}}, \bibinfo
  {author} {\bibfnamefont {L.~O.}\ \bibnamefont {Silva}}, \ and\ \bibinfo
  {author} {\bibfnamefont {W.~B.}\ \bibnamefont {Mori}},\ }\href {\doibase
  10.1086/427921} {\bibfield  {journal} {\bibinfo  {journal} {Astrophys. J.}\
  }\textbf {\bibinfo {volume} {618}},\ \bibinfo {pages} {L75} (\bibinfo {year}
  {2005})}\BibitemShut {NoStop}%
\bibitem [{\citenamefont {Cary}\ \emph {et~al.}(1981)\citenamefont {Cary},
  \citenamefont {Thode}, \citenamefont {Lemons}, \citenamefont {Jones},\ and\
  \citenamefont {Mostrom}}]{Cary1981}%
  \BibitemOpen
  \bibfield  {author} {\bibinfo {author} {\bibfnamefont {J.~R.}\ \bibnamefont
  {Cary}}, \bibinfo {author} {\bibfnamefont {L.~E.}\ \bibnamefont {Thode}},
  \bibinfo {author} {\bibfnamefont {D.~S.}\ \bibnamefont {Lemons}}, \bibinfo
  {author} {\bibfnamefont {M.~E.}\ \bibnamefont {Jones}}, \ and\ \bibinfo
  {author} {\bibfnamefont {M.~A.}\ \bibnamefont {Mostrom}},\ }\href {\doibase
  10.1063/1.863262} {\bibfield  {journal} {\bibinfo  {journal} {Phys. Fluids}\
  }\textbf {\bibinfo {volume} {24}},\ \bibinfo {pages} {1818} (\bibinfo {year}
  {1981})}\BibitemShut {NoStop}%
\bibitem [{\citenamefont {Stockem}, \citenamefont {Dieckmann},\ and\
  \citenamefont {Schlickeiser}(2008)}]{Stockem2008}%
  \BibitemOpen
  \bibfield  {author} {\bibinfo {author} {\bibfnamefont {A.}~\bibnamefont
  {Stockem}}, \bibinfo {author} {\bibfnamefont {M.~E.}\ \bibnamefont
  {Dieckmann}}, \ and\ \bibinfo {author} {\bibfnamefont {R.}~\bibnamefont
  {Schlickeiser}},\ }\href {\doibase 10.1088/0741-3335/50/2/025002} {\bibfield
  {journal} {\bibinfo  {journal} {Plasma Phys. Controll. Fusion}\ }\textbf
  {\bibinfo {volume} {50}},\ \bibinfo {pages} {025002} (\bibinfo {year}
  {2008})}\BibitemShut {NoStop}%
\bibitem [{\citenamefont {Bret}(2009)}]{Bret2009}%
  \BibitemOpen
  \bibfield  {author} {\bibinfo {author} {\bibfnamefont {A.}~\bibnamefont
  {Bret}},\ }\href {\doibase 10.1088/0004-637X/699/2/990} {\bibfield  {journal}
  {\bibinfo  {journal} {Astrophys. J.}\ }\textbf {\bibinfo {volume} {699}},\
  \bibinfo {pages} {990} (\bibinfo {year} {2009})}\BibitemShut {NoStop}%
\bibitem [{\citenamefont {Grassi}\ \emph {et~al.}(2017)\citenamefont {Grassi},
  \citenamefont {Grech}, \citenamefont {Amiranoff}, \citenamefont {Pegoraro},
  \citenamefont {Macchi},\ and\ \citenamefont {Riconda}}]{Grassi2017}%
  \BibitemOpen
  \bibfield  {author} {\bibinfo {author} {\bibfnamefont {A.}~\bibnamefont
  {Grassi}}, \bibinfo {author} {\bibfnamefont {M.}~\bibnamefont {Grech}},
  \bibinfo {author} {\bibfnamefont {F.}~\bibnamefont {Amiranoff}}, \bibinfo
  {author} {\bibfnamefont {F.}~\bibnamefont {Pegoraro}}, \bibinfo {author}
  {\bibfnamefont {A.}~\bibnamefont {Macchi}}, \ and\ \bibinfo {author}
  {\bibfnamefont {C.}~\bibnamefont {Riconda}},\ }\href {\doibase
  10.1103/PhysRevE.95.023203} {\bibfield  {journal} {\bibinfo  {journal} {Phys.
  Rev. E}\ }\textbf {\bibinfo {volume} {95}},\ \bibinfo {pages} {023203}
  (\bibinfo {year} {2017})}\BibitemShut {NoStop}%
\bibitem [{\citenamefont {Pelletier}\ \emph {et~al.}(2019)\citenamefont
  {Pelletier}, \citenamefont {Gremillet}, \citenamefont {Vanthieghem},\ and\
  \citenamefont {Lemoine}}]{Pelletier2019}%
  \BibitemOpen
  \bibfield  {author} {\bibinfo {author} {\bibfnamefont {G.}~\bibnamefont
  {Pelletier}}, \bibinfo {author} {\bibfnamefont {L.}~\bibnamefont
  {Gremillet}}, \bibinfo {author} {\bibfnamefont {A.}~\bibnamefont
  {Vanthieghem}}, \ and\ \bibinfo {author} {\bibfnamefont {M.}~\bibnamefont
  {Lemoine}},\ }\href {\doibase 10.1103/PhysRevE.100.013205} {\bibfield
  {journal} {\bibinfo  {journal} {Phys. Rev. E}\ }\textbf {\bibinfo {volume}
  {100}},\ \bibinfo {pages} {013205} (\bibinfo {year} {2019})}\BibitemShut
  {NoStop}%
\bibitem [{\citenamefont {Dieckmann}\ \emph {et~al.}(2007)\citenamefont
  {Dieckmann}, \citenamefont {Lerche}, \citenamefont {Shukla},\ and\
  \citenamefont {Drury}}]{Dieckmann2007}%
  \BibitemOpen
  \bibfield  {author} {\bibinfo {author} {\bibfnamefont {M.~E.}\ \bibnamefont
  {Dieckmann}}, \bibinfo {author} {\bibfnamefont {I.}~\bibnamefont {Lerche}},
  \bibinfo {author} {\bibfnamefont {P.~K.}\ \bibnamefont {Shukla}}, \ and\
  \bibinfo {author} {\bibfnamefont {L.~O.~C.}\ \bibnamefont {Drury}},\ }\href
  {\doibase 10.1088/1367-2630/9/1/010} {\bibfield  {journal} {\bibinfo
  {journal} {New J. Phys.}\ }\textbf {\bibinfo {volume} {9}},\ \bibinfo {pages}
  {10} (\bibinfo {year} {2007})}\BibitemShut {NoStop}%
\bibitem [{\citenamefont {{JÃ¶nsson}}\ \emph {et~al.}(2019)\citenamefont
  {{JÃ¶nsson}}, \citenamefont {{Steneteg}}, \citenamefont {{SundÃ©n}},
  \citenamefont {{Englund}}, \citenamefont {{Kottravel}}, \citenamefont
  {{Falk}}, \citenamefont {{Ynnerman}}, \citenamefont {{Hotz}},\ and\
  \citenamefont {{Ropinski}}}]{Inviwo}%
  \BibitemOpen
  \bibfield  {author} {\bibinfo {author} {\bibfnamefont {D.}~\bibnamefont
  {{JÃ¶nsson}}}, \bibinfo {author} {\bibfnamefont {P.}~\bibnamefont
  {{Steneteg}}}, \bibinfo {author} {\bibfnamefont {E.}~\bibnamefont
  {{SundÃ©n}}}, \bibinfo {author} {\bibfnamefont {R.}~\bibnamefont
  {{Englund}}}, \bibinfo {author} {\bibfnamefont {S.}~\bibnamefont
  {{Kottravel}}}, \bibinfo {author} {\bibfnamefont {M.}~\bibnamefont {{Falk}}},
  \bibinfo {author} {\bibfnamefont {A.}~\bibnamefont {{Ynnerman}}}, \bibinfo
  {author} {\bibfnamefont {I.}~\bibnamefont {{Hotz}}}, \ and\ \bibinfo {author}
  {\bibfnamefont {T.}~\bibnamefont {{Ropinski}}},\ }\href {\doibase
  10.1109/TVCG.2019.2920639} {\bibfield  {journal} {\bibinfo  {journal} {IEEE
  Transactions on Visualization and Computer Graphics}\ ,\ \bibinfo {pages}
  {1}} (\bibinfo {year} {2019})}\BibitemShut {NoStop}%
\end{thebibliography}%

\end{document}